\documentclass{article}
\usepackage{graphicx}
\usepackage{latexsym}
\usepackage[cp1250]{inputenc}
\newcommand{\M}{{\cal M}}
\newcommand{\PT}{{\cal P}{\cal T}}
\newcommand{\hol}{{\cal O}}

\newcommand{\Q}{{\cal Q}}
\newcommand{\PP}{{\cal P}}
\newcommand{\A}{{\cal A}}
\newcommand{\eQ}{e^{{\cal Q}}}

\newcommand{\h}{\hspace{.5 em}}
\newcommand{\vs}{\vspace{1 ex}}
\newcommand{\bdm}{\begin{displaymath}}
\newcommand{\edm}{\end{displaymath}}
\newcommand{\bi}{\begin{itemize}}
\newcommand{\ei}{\end{itemize}}
\newcommand{\benum}{\begin{enumerate}}
\newcommand{\eenum}{\end{enumerate}}
\newcommand{\be}{\begin{equation}}
\newcommand{\ee}{\end{equation}}
\newcommand{\bea}{\begin{eqnarray}}
\newcommand{\eea}{\end{eqnarray}}
\newcommand{\btabular}{\begin{tabular}}
\newcommand{\etabular}{\end{tabular}}
\newcommand{\beas}{\begin{eqnarray*}}
\newcommand{\eeas}{\end{eqnarray*}}

\newtheorem{defin}{Definition}[section]
\newcommand{\bdefin}{\begin{defin}}
\newcommand{\edefin}{\end{defin}}
\newtheorem{theorem}{Theorem}[section]
\newcommand{\bthrm}{\begin{theorem}}
\newcommand{\ethrm}{\end{theorem}}
\newtheorem{proposition}[defin]{Proposition}
\newcommand{\bproposition}{\begin{proposition}}
\newcommand{\eproposition}{\end{proposition}}
\newtheorem{corollary}[defin]{Corollary}
\newcommand{\bcorollary}{\begin{corollary}}
\newcommand{\ecorollary}{\end{corollary}}
\newtheorem{rem}[defin]{Remark}
\newcommand{\brem}{\begin{rem}}
\newcommand{\erem}{\end{rem}}
\newtheorem{Lem}[defin]{Lemma}
\newcommand{\bL}{\begin{Lem}}
\newcommand{\eL}{\end{Lem}}
\newtheorem{exam}[defin]{Example}
\newcommand{\bexam}{\begin{exam}}
\newcommand{\eexam}{\end{exam}}
\begin{document}
\begin{center}
\Large $\ast$-SDYM Fields and Heavenly Spaces. \normalsize\\
\mbox{ }\\
\large I. $\ast$-SDYM equations as an integrable system. \normalsize\\

\mbox{ }\\

Sebastian Formański$^{1}$\footnote{sforman@p.lodz.pl}
and Maciej Przanowski$^{1,2}$\footnote{przan@fis.cinvestav.mx} \\
\mbox{ }\\
$^{1}$Institute of Physics,Technical University of Lodz,\\
W\'{o}lcza\'{n}ska 219. 93-005 Łódź, Poland\\
\mbox{ }\\
$^{2}$ Departamento de Fisica\\
Centro de Investigaciones y de Estudios Avanzados del IPN\\
Apartado Postal 14-740, 07000 Mexico D.F. Mexico.\\ 
\end{center}
\paragraph{Abstract.}
It is shown that the self-dual Yang-Mills (SDYM) equations  for the $\ast$-bracket Lie algebra on a heavenly 
space can be reduced to one equation (the \it master equation\rm). Two hierarchies of conservation 
laws for this equation are constructed. Then the twistor transform and a solution to the 
Riemann-Hilbert problem are given.

\vspace {3 ex}

\section*{Introduction} 
It turns out that many nonlinear integrable systems are reductions of SDYM equations 
(e.g. see Mason and Woodhouse 1996). The statement that all integrable system of mathematical
physics are some reductions of SDYM equations is known as \it Ward's conjecture  \rm (Ward 1985).
The twistor construction for SDYM system is in a sense inherited by the reduced system. 
There are however, exceptional cases which do not fit into this scheme for a finite dimensional 
structure group (Mason and Woodhouse 1996). 

An extension of the Lie algebra of SDYM equations to infinite dimensional algebra 
of Hamiltonian vector fields provides description of heavenly spaces of complex general 
relativity (Mason 1989). The nonlinear graviton construction (Penrose 1976, Penrose and Ward 1980, 
Mason and Woodhouse 1996) proves that the heavenly equations constitute an integrable system. Thus 
the idea arose, that ${\cal H}$-space might be an universal integrable system (Mason 1990). 
However, the 
reduction of the algebra of Hamiltonian vector fields over a symplectic manifold $\Sigma^{2}$, 
$sdiff(\Sigma^{2})$ to finite dimensional algebras such as $su(N)$ does not exist for $N>2$. 

Consequently, it seems that Ward's conjecture should be extended to the algebras that include 
all finite dimensional Lie algebras $sl(N,C)$ as well as the algebra of hamiltonian vector fields.
This is the point where \it deformation quantization \rm enters into the theory of
integrable systems.

The idea of deformation quantization, introduced by Bayen, Flato, Fronsdal, Lichnerowicz and 
Strenhaimer in (Bayen et al. 1978) is to consider a deformed algebra of smooth functions on a 
classical phase space. The introduced associative $\ast$-product of two functions $f,g$ is a 
formal power series in deformation parameter $\hbar$,  $f\ast g=\sum\hbar^{k}\Delta_{k}(f,g)$, \h 
$k\geq 0$. The $\ast$-product is assumed to satisfy the following axioms:  

$\bullet$ it is local i.e. $\Delta_{k}(f,g)$ depends only on $f, g$ and 
partial derivatives of $f, g$ of rank not greater than $k$,
 
$\bullet$ it is a deformation of Poisson algebra i.e. 
$\Delta_{1}(f,g)-\Delta_{1}(g,f)=i\{f,g\}_{\mbox{\footnotesize{Poisson}}}$.

One can prove that such a product exists on any symplectic \linebreak
(De Wilde and Lecomte 1983, Fedosov 1994) or even Poisson manifold (Kontsevich 1997).

It seems to be useful to consider integrable (quantum) deformations of integrable 
systems (Kupershmidt 1990, Strachan 1992 and 1997, Takasaki 1994). It is so, because the Moyal bracket 
algebra can be reduced to all $su(N)$ algebras (Fairlie et al. 1990). In a natural way the Poisson 
algebra is embedded in deformed algebra. This suggests that SDYM equations for $\ast$-bracket 
Lie algebra ($\ast$-SDYM equations) are reducible to $su(N)$-SDYM equations as well as to heavenly 
equations. This is the problem which we intend to consider in the present  and the next papers. 
The present paper is devoted mainly to formal problem of integrability of $\ast$-SDYM equations. In order to 
make our results more general we deal with an arbitrary $\ast$-product, and the Yang-Mills fields
are defined on 4-dimensional heavenly space. 

It turns out that to have the formalism sufficiently general one needs to deal with  
formal power series containing all negative powers of deformation parameter $\hbar$,  
in particular with the power series of the form $\exp[\frac{1}{i\hbar}A]$. 
As is well known (Fedosov 1996) such power series are well defined only for 
some special $A$. To ensure the existence of exponent for wide class of $A$ we introduce
new formal parameter $t$ (\it convergence parameter\rm ). It is obvious that in applications
only such series will be used that are convergent with respect to the \mbox{parameter $t$.}

Our paper is organized as follows. In section 1 we give some fundamental definitions and 
properties of formal power series. Then we obtain a group $\eQ$ of formal power series
suitable for construction of respective gauge theory. The $\ast$-SDYM  equations on 
K\H{a}hler manifold in the case of heavenly space are reduced to one equation called 
\it master equation \rm (ME) (\ref{ME in H}). In section 2 we find two collections of 
conserved charges (\ref{defin of R}) and (\ref{recursion 2}). As is pointed out the 
collection (\ref{recursion 2}) is characteristic for any SDYM system and (\ref{defin of R}) 
is a generalization of hidden symmetries of heavenly or SDYM equations. 
We obtain two Lax pairs and the forward Penrose-Ward transform for ME. 
Dressing operator connecting those two pairs, and finally, the algebra of hidden symmetries 
are given. Section 3 is devoted to the solution of Riemann-Hilbert problem. We define 
the homogeneous Hilbert problem and we show the existence of the solution of this problem 
for the formal power series group $\eQ$ (\it Birkhoff's factorization theorem\rm ). Then the 
inverse Penrose-Ward transform is considered. Concluding remarks (section 4) close the paper.

Some applications of master equation (ME) in the theory of inegrable systems and complex
relativity will be presented in the forthcoming paper. In that paper a sequence of $su(N)$ 
chiral fields tending to the heavenly space for $N\rightarrow\infty$ has been constructed. 
It has been also shown that any analytic solution of $su(N)$ SDYM  equations can be obtained 
from some solution of $\ast$-SDYM equations.
\section{Formal power series and $\ast$-SDYM equations.}
At the beginning of this section we briefly summarize the basic definitions and theorems 
concerning formal power series. We define algebra, group and adjoint action. For more details 
see (MacLane 1939, Neumann 1949, Jacobson 1980, Ruiz 1993).

The \it ordered abealian group \rm is a pair $((G,+),P)$, where $(G,+)$ is an abealian group, 
$P$ is a subset of $G$ such that:

$\bullet$ $0\notin P\, ,$\h $P\cap -P=\emptyset$ \h ($0$ is the neutral element of $(G,+)$)

$\bullet$ $\forall g,h\in P\, , \h g+h\in P$

$\bullet$ $G=-P\cup\{0\}\cup P$.

\noindent We call $P$ the subset of positive elements. It allows one to order elements of the 
group i.e. if $g,h\in G$ we say that $g$ \it is less than \rm $h$ and denote $g<h$ 
if and only if $h-g\in P$.

Let $((G,+),P)$ be an ordered group and $\mbox{\boldmath{$K$}}$ a vector space. \it Formal 
power series over $G$ with coefficients in $\mbox{\boldmath{$K$}}$ \rm is a map 
$a:G\rightarrow\mbox{\boldmath{$K$}}$, such that its support 
$\mbox{supp}\, a=\{g\in G\, : \h a(g)\neq 0\}$ has the least element.

The formal power series $a$ will be written in the following form 
\bdm
a=\sum_{g\in G}a_{g}\hbar^{g}\h\h\mbox{where}\h\h a_{g}=a(g),\h\h \hbar\mbox{-parameter.}
\edm
The set of all formal power series over $G$ with coefficients in $\mbox{\boldmath{$K$}}$ 
will be denoted by  $\mbox{\boldmath{$K$}}((\hbar^{G}))$. It is a vector space over 
complex field, with addition and  multiplication by scalar are defined pointwise by
\beas
a+b=\sum_{g\in G}(a_{g}+b_{g})\hbar^{g}\\
\forall\alpha\in\mbox{\boldmath{$C$}}\h\h \alpha a=\sum_{g\in G}\alpha a_{g}\hbar^{g}.
\eeas
Moreover if the pair $(\mbox{\boldmath{$K$}},\circ)$ is an algebra then the multiplications 
of series is defined
\bdm
ab=\sum_{g\in G}(\sum_{h\in G}a_{h}\circ b_{g-h})\hbar^{g}.
\edm
This multiplication is well defined, as the support of each series has the least element, so
$\forall g\in G$ the number of elements $a_{h}\circ b_{g-h}$, $h\in G$ is finite.

In the case when $(G,+)$ is a group $(Z,+)$ we will write $\mbox{\boldmath{$K$}}((\hbar))$. 
Moreover $\mbox{\boldmath{$K$}}[[\hbar]]:=\{a\in\mbox{\boldmath{$K$}}((\hbar))\, ,\, 
a_{g}=0 \h \forall g\in -P\}$.

\vspace{2 ex}

According to Fedosov's works (Fedosov 1994, Fedosov 1996) the pair 
$(\hol(\Sigma^{2n})[[\hbar]],\ast)$ constitutes an algebra. 
$\hol(\Sigma^{2n})[[\hbar]]$ denotes linear space of formal power 
series with coefficients being 
holomorphic functions over symplectic manifold $(\Sigma^{2n}, \omega )$.
$\ast$ is an associative and noncommutative multiplication \linebreak
\mbox{$\ast: \hol(\Sigma^{2n})[[\hbar]]\times\hol(\Sigma^{2n})[[\hbar]]\rightarrow
\hol(\Sigma^{2n})[[\hbar]]$.} The $\ast$-product considered is a \it closed 
$\ast$-product \rm i.e. the trace
$\mbox{tr}(f\ast g):=\int\frac{\omega^{n}}{n!}f\ast g $
has the property  $\mbox{tr}(f\ast g)=\mbox{tr}(g\ast f)$ (Connes et al 1992, 
Omori et al 1992, Fedosov 1996) 

We can define the Lie algebra \mbox{$(\hol(\Sigma^{2n})[[\hbar]] , \{\, ,\,\} )$}
based on $\ast$-product
\bdm
\forall a,b\in\hol(\Sigma^{2n})[[\hbar]] \h\h\h \{ a, b\} =\frac{1}{i\hbar}(a\ast b-b\ast a).
\edm

Our aim is to construct gauge theory. The fundamental object is the gauge group. 
The group element appears as an exponent of the element of Lie algebra. In finite 
dimensional case the exponent of left-invariant vector fields is the maximal integral 
curve,  a 1-parameter subgroup of the Lie group. 

In the case of the $\ast$-algebra  taking an exponent is possible only for some 
special vectors (Compare Fedosov 1996, Asakawa and Kishimoto 2000). Such a group will not be 
general enough to define gauge transformation.

In order to make superpositions of formal power series well defined, we need to 
introduce another parameter. So we would consider formal power series over $(Z,+)$ 
with coefficients in a space $\hol(\Sigma^{2n})((\hbar))$. In a space 
$\hol(\Sigma^{2n})((\hbar, t))$ of all such series we consider a subspace $\A$ 
\bdm
\A=\{\h A\in \hol(\Sigma^{2n})((\hbar, t))\h , \h
A=\sum_{m=0}^{\infty}\sum_{k=-m}^{\infty} t^{m}\hbar^{k}A_{m,k}(x)\h \}.
\edm
The star product $\ast$ defined on $\hol(\Sigma^{2n})[[\hbar]]$ can be extended to
$\A$ (we use the same symbol)
\beas
\forall A,B\in \A,\h\h
A=\sum_{m_{1}=0}^{\infty}\sum_{k_{1}=-m_{1}}^{\infty} t^{m_{1}}\hbar^{k_{1}}A_{m_{1},k_{1}}(x)\, ,\h
B=\sum_{m_{2}=0}^{\infty}\sum_{k_{2}=-m_{2}}^{\infty} t^{m_{2}}\hbar^{k_{2}}B_{m_{2},k_{2}}(x)\\
A\ast B=\sum_{m_{1},m_{2}=0}^{\infty} t^{m_{1}+m_{2}}\hbar^{-(m_{1}+m_{2})}
\sum_{k_{1}=-m_{1},k_{2}=-m_{2}}^{\infty}(\hbar^{k_{1}+m_{1}}A_{m_{1},k_{1}}(x)\ast
\hbar^{k_{2}+m_{2}}B_{m_{2},k_{2}}(x)).
\eeas
The algebra $(\A,\ast)$ is called \it formal $\ast$-algebra\rm .

\vs

Let $A\in \A$. The element $A(0)$ i.e. the one which stands at $t^{0}$
will be denoted $\phi(A)$ and called \it the free element \rm
\bdm
\forall A\in\A\h\h\h A=\phi(A)+\sum_{m=1}^{\infty}\sum_{k=-m}^{\infty}t^{m}\hbar^{k}A_{m,k},
\h\h\mbox{where}\h\h\phi(A)=\sum_{k=0}^{\infty}\hbar^{k}A_{0,k}.
\edm
The family ${\cal N}$ of formal power series belonging to a formal $\ast$-algebra $\A$, 
\mbox{${\cal N}=\{A_{\delta}\in \A\, ,\h \delta\in\Omega\}$} is called \it $t$-locally
finite \rm if for each natural $m$ the number of formal power series of this family 
having non-zero element at $t^{m}$ is finite.

Then for each t-locally finite family and any family of complex number 
$\{a_{\delta}\in\mbox{\boldmath$C$}\, ,\h\delta\in\Omega\}$  the sum 
$\sum_{\delta\in\Omega}a_{\delta}A_{\delta}$ is well
defined
\bi
\item Let $A\in \A$ and let free element $\phi(A)=0$. Then the family 
$\{A^{n}, n=1,2,...\}$,
where $A^{n}\equiv A\ast A^{n-1}=A^{n-1}\ast A$ is  $t$-locally finite.

\item Let $f(z)=\sum_{n=0}^{\infty}a_{n}z^{n}$ be a complex power series of one
variable and $A\in \A$ with  $\phi(A)=0$. 
We define $f(A)=\sum_{n=0}^{\infty}a_{n}A^{n}$ (where $A^{n}$ is as above).

\item If $f_{1}(z)=\sum_{n=0}^{\infty}a_{n}z^{n}$ and 
$f_{2}(z)=\sum_{m=0}^{\infty}b_{m}z^{m}$ 
are two formal power series and $A\in \A$ with $\phi(A)=0$, then
\bdm
(f_{1}\cdot f_{2})(A)=(f_{2}\cdot f_{1})(A)=f_{1}(A)\ast f_{2}(A)
\edm
This follows from the fact that $\A$ is an algebra with 
$A^n \ast A^m=A^m \ast A^n=A^{n+m}$, and multiplication of coefficients 
$a_{n}, b_{m}$ is commutative.
\item For each $A\in \A$ such that $\phi(A)=1$ there exists 
inversion of $A$ i.e. $X\in \A$, $A\ast X=X\ast A=1$. 

Indeed, let $f_{1}(z):=(z+1)^{-1}=\sum_{n=0}^{\infty}(-1)^{n}z^{n}$.
Let $f_{2}(z):=z+1$, then $(f_{1}\cdot f_{2})(z)=1$.
If one defines  $X=f_{1}(A-1)$,
\beas
1&=&(f_{1}\cdot f_{2})(A-1)=f_{1}(A-1)\ast f_{2}(A-1)=f_{1}(A-1)\ast A\\
1&=&(f_{2}\cdot f_{1})(A-1)=f_{2}(A-1)\ast f_{1}(A-1)=A\ast f_{1}(A-1)
\eeas
In what follows we will write $A^{-1}:=X$.
\ei
Remarks

\bi
\item $\phi(A^{-1})=1$
\item The set $\{A\in\A\, ,\h \phi(A)=1\}$ with $\ast$-product form a group, it is 
a subgroup of invertible elements.
\ei
Let $f_{1}(z)=\sum_{n=0}^{\infty}a_{n}z^{n}$, $f_{2}(z)=\sum_{m=1}^{\infty}b_{m}z^{m}$ 
and $A\in \A$ with $\phi(A)=0$. 
\bi
\item Then superposition of series $f(A)=f_{1}(f_{2}(A))$ is well defined.
\ei
This follows from the fact that the family $\{[f_{2}(A)]^{n}\, , \, n=0,1,2,...\}$
is \mbox{t-locally} finite.

\bcorollary\label{wniosek eksponent}\rm For each $A\in\A$ with $\phi(A)=0$, \h
$e^{A}:=\sum_{n=0}^{\infty}\frac{A^{n}}{n!}$
is an element of the group with the free element equal $1$. On the other hand each element 
of this group is an exponent. The inversion map is given by
$A=\sum_{n=1}^{\infty}\frac{(-1)^{n+1}}{n}(e^{A}-1)^{n}$.
\ecorollary

\noindent In what follows $\Q$ is subalgebra of formal power series with free element
equal to zero
\bdm
\Q:=\{\h A\in \A \h ,\h A=\sum_{m=1}^{\infty}\sum_{k=-m}^{\infty}t^{m}\hbar^{k}A_{m,k}(x)\h \}
\edm
$\eQ$ is  a group of formal power series with free element equal $1$ 
\bdm
\eQ:=\{\h a\in \A \h ,\h a=1+\sum_{m=1}^{\infty}\sum_{k=-m}^{\infty}t^{m}\hbar^{k}a_{m,k}(x)\h \}
\edm
From Corollary \ref{wniosek eksponent} the algebra $\Q$ with Lie bracket 
$\forall A,B\in\Q\h\h \{A,B\}:=\frac{1}{i\hbar}(A\ast B-B\ast A)$
will be called the Lie algebra of the group $\eQ$. 

It is worth noting here that we do not consider differential structure on the group
$\eQ$, so this is not a Lie group. Apart from that,  for $A,B\in \Q$ one has
\bdm
\frac{1}{i\hbar}\frac{d}{d\varepsilon}|_{\varepsilon=0}\, 
( e^{-\sqrt{\varepsilon}A}\ast e^{-\sqrt{\varepsilon}B}\ast
e^{\sqrt{\varepsilon}A}\ast e^{\sqrt{\varepsilon}B})=\{A,B\}
\edm
This justifies our notation.

From Corollary \ref{wniosek eksponent} $\forall a\in\eQ$ there exists$\tilde{A}\in \Q$, 
such that $a=\exp(\tilde{A})$. For traditional reasons we will write
\be
\label{umowa wykladnik}
a=e^{\frac{1}{i\hbar}A}\h\h\mbox{where}\h\h\tilde{A}=
\frac{1}{i\hbar}\sum_{m=1}^{\infty}\sum_{k=-m+1}^{\infty}t^{m}\hbar^{k}A_{m,k}(X)=:
\frac{1}{i\hbar}A
\ee

\noindent For our purpose it is important to consider the following left actions of $\eQ$ 
on the algebra $\A$
\bea
\label{action1}
\psi:\eQ\times \A  \rightarrow \A\h &,&\h\h\psi(a,f):= a\ast f\\
\label{action2}
\phi:\eQ\times \A  \rightarrow \A\h &,&\h\h\phi(a,f):= f\ast a^{-1}
\eea
and the adjoint representation
\be
\label{adjointaction}
Ad:\eQ\times \A \rightarrow \A\h ,\h\h Ad(a,f):= a\ast f\ast a^{-1}.
\ee
Acording to (\ref{umowa wykladnik}) the adjoint representation can be written 
in the following form
\be
\label{dzialanie adjoint}
a\ast f\ast a^{-1}=f+\sum_{l=1}^{\infty}\frac{1}{l!}
\underbrace{\{A,...,\{A}_{l-\mbox{\tiny razy}},f\}\cdots\}.
\ee
Each of the actions do not change the free element. See Asakawa and Kishimoto (2000).
\paragraph{Bundle of formal $\ast$-Algebras.}
Let us consider four dimensional complexified K\H{a}hler manifold $\M$\footnote{
The coordinates on $\M$ are denoted by \mbox{$(w, z, \tilde{w}, \tilde{z})$} we use 
also the following abbreviations $z^{\alpha}=\{w, z\}$, 
$z^{\tilde{\alpha}}=\{\tilde{w}, \tilde{z}\}$ and 
$z^{i}=\{w, z, \tilde{w}, \tilde{z}\}$. 
$\M$ is hermitian manifold i.e. is equipped with holomorphic nondegenerate
metric $ds^{2}=2g_{\alpha\tilde{\beta}}dz^{\alpha}\otimes_{s}d z^{\tilde{\beta}}$. This 
reduces the allowed transformations to the ones which preserve the foliation $w=const.$, 
$z=const.$ as well as $\tilde{w}=const,$, $\tilde{z}=const.$. Each 1-form
$\sigma\in\Lambda^{1}\M$ can be decomposed to a sum $\sigma=\sigma_{(1,0)}+\sigma_{(0,1)}$\h 
where $\sigma_{(1,0)}=\sigma_{w}dw+\sigma_{z}dz$,\h and 
$\sigma_{(0,1)}=\sigma_{\tilde{w}}d\tilde{w}+\sigma_{\tilde{z}}d\tilde{z}$.
Analogously, the exterior derivative $d$ is a sum of two Dolbeaut operators 
$d=\partial +\tilde{\partial}$ where 
\bdm
\partial=dw\wedge\partial_{w}+dz\wedge\partial_{z}\h ,\h\h
\tilde{\partial}=d\tilde{w}\wedge\partial_{\tilde{w}}+
d\tilde{z}\wedge\partial_{\tilde{z}}.
\edm
\h The \it K\H{a}hler form\rm is a 2-form of the type $(1,1)$, given by 
$\Omega=g_{\alpha\tilde{\beta}}\, dz^{\alpha}\wedge d\tilde{z}^{\beta}$.
For \it complexified K\H{a}hler manifold \rm the K\H{a}hler form is closed  $d\Omega=0$. 
Locally this means that $\Omega=\partial\tilde{\partial}{\cal K}$ for some complex 
function ${\cal K}={\cal K}(w,z,\tilde{w},\tilde{z})$ called \it K\H{a}hler potential\rm .
The K\H{a}hler form $\Omega=g_{\alpha\tilde{\beta}}dz^{\alpha}\wedge dz^{\tilde{\beta}}$
gives rise to the \it volume element \rm 
$\mbox{\boldmath$\nu$}:=\frac{1}{2}\Omega\wedge \Omega\, =\, g\, dw\wedge d\tilde{w}
\wedge dz\wedge d\tilde{z}$ where 
$g=\det(g_{\alpha\tilde{\beta}})={\cal K},_{w\tilde{w}}{\cal K},_{z\tilde{z}}-
{\cal K},_{w\tilde{z}}{\cal K},_{z\tilde{w}}$. 
Under the Hodge duality the following 2-forms constitute the basis
of anti-self-dual forms
\bdm
\Sigma^{\stackrel{.}{1}\stackrel{.}{1}}:=d\tilde{x}\wedge d\tilde{y},
\hspace{1 ex}
\Sigma^{\stackrel{.}{1}\stackrel{.}{2}}:=\Omega,\hspace{1 ex}, 
\Sigma^{\stackrel{.}{2}\stackrel{.}{2}}:=dx\wedge dy
\edm
(See Pleba\'{n}ski 1975, Flaherty 1976, Ko et al. 1981, Mason \& Woodhouse 1996 for more details.)}.
We will consider functions and tensors on $\M$ wich takes values in 
formal \mbox{$\ast$-algebra $\A$.}

\vs

Let $\PP(\M,\eQ)$ denote a trivial principle bundle $\pi_{\small{\PP}}:\PP\rightarrow\M$. 
In a total space $\PP$ the structure group acts to the right in each fibre 
$\PP\times\eQ\ni(u,c)\mapsto u\ast c\in\PP$. The global sections are then
\bdm
\sigma=1+\sum_{m=1}^{\infty}\sum_{k=-m}^{\infty}t^{m}\hbar^{k}
\sigma_{m,k}(x, z^{i})  \h\h\mbox{gdzie}\h\h
\sigma_{m,k}(x, z^{i})\in \hol(\Sigma^{2n}\times\M)
\edm
Each representation of the group $\eQ$ in algebra  $\A$ defined by (\ref{action1}), 
(\ref{action2}), (\ref{adjointaction})  allows one to define an associated bundle with 
tipical fibre $\A$. In the Cartesian product $\PP\times\A$ we introduce the following 
equivalence relations
\beas
(u,v)\sim_{\psi} (u',v')\h & \Leftrightarrow &  [\pi_{\PP}(u)=\pi_{\PP}(u'),\h
\exists c\in \eQ, \h u'=u\ast c\, ,\h v'=c^{-1}\ast v\, ],\\
(u,v)\sim_{\phi} (u',v')\h & \Leftrightarrow & [\pi_{\PP}(u)=\pi_{\PP}(u'),\h
\exists c\in \eQ, \h u'=u\ast c\, ,\h v'=v\ast c\, ],\\
(u,v)\sim_{Ad} (u',v')\h & \Leftrightarrow & [\pi_{\PP}(u)=\pi_{\PP}(u'),\h
\exists c\in \eQ, \h u'=u\ast c\, ,\h v'=c^{-1}\ast v\ast c\, ].
\eeas

\bdefin\rm
\mbox{                                                                                       }
\bi
\item\it The formal $\ast$-algebra bundle $E(\M, \A)$ over $\M$ \rm is a set 
of equivalence classes of Cartesian product\h $\PP\times\A$\h in relation 
$\sim_{\psi}$ i.e. the total space is $E:=\PP\times \A /\sim_{\psi}$.
\ei

\noindent Analogously
\bi
\item \it The bundle \rm $E':=\PP\times\A/\sim_{\phi}$.
\item \it The adjoint bundle \rm \h $adj(E):=\PP\times\A/\sim_{Ad}$.
\ei
The map $\pi_{\PP}$ induces the maps $\pi_{\small{E}}:E\rightarrow \M$, 
$\pi_{\small{E'}}:E'\rightarrow \M$,
$\pi_{\small{adj(E)}}:adj(E)\rightarrow \M$.
These maps send the equivalence class $[(u, v)]$ to 
$\pi_{\PP}(u)\in\M$ what makes that $E, E', adj(E)$  are bundles.
\edefin

For each section $\sigma:\M\rightarrow\PP$ of principal bundle, the section
$f:\M\rightarrow E$ of associated bundle induces map
$\tilde{f}_{\sigma}:\PP\rightarrow\PP\times\A$ such that the following diagram 
is commutative

\setlength{\unitlength}{5ex}
\begin{picture}(9,5)
\put(1,1){$\M$}
\put(1.8,1.1){\vector(1,0){3.6}}
\put(3.2,1.25){$f$}
\put(6.3,1){E}
\put(1.3,1.5){\vector(0,1){2}}
\put(0.5,2.5){$\sigma$}
\put(1.1,4){$\PP$}
\put(1.8,4.1){\vector(1,0){3.6}}
\put(3.2,4.25){$\tilde{f}_{\sigma}$}
\put(5.8,4){$\PP\times\A$}
\put(6.5,3.5){\vector(0,-1){2}}
\put(6.8,2.5){$\ell$}
\end{picture}


\noindent where $\ell$ is the canonical map $\ell(u,v)=[(u,v)]$.

\noindent The map $\tilde{f}_{\sigma}$ defines another map
$f_{\sigma}:\M\rightarrow\A$ by 
\mbox{$f_{\sigma}:=pr_{\A}\circ\tilde{f}_{\sigma}\circ\sigma$.} It is called
\it the representation of the section $f:\M\rightarrow E$ with respect to the section \rm 
$\sigma:\M\rightarrow\PP$.

\vspace{1 ex}

Higher external powers of the bundle$E$, $E^{k}:=E\otimes\Lambda^{k}\M$
k=1,2,3,4 form an algebra of forms with values in $E$
\bdm
E\otimes\Lambda\M:=\bigoplus_{k=1}^{4} E\otimes\Lambda^{k}\M\, ,
\edm
with external product (in terms of representationes)
\bdm
\mbox{\boldmath$\omega$}\wedge\mbox{\boldmath$\nu$}:=
\omega_{i_{1}...i_{k}}\ast \nu_{j_{1}...j_{l}}
dx^{i_{1}}\wedge ...\wedge dx^{i_{k}}\wedge dx^{j_{1}}\wedge
...\wedge dx^{j_{l}}.
\edm
This allows to define \it the bracket of forms \rm
\beas
\{\mbox{\boldmath$\omega$},\mbox{\boldmath$\nu$}\}
&:=&\frac{1}{i\hbar}(\,
\mbox{\boldmath$\omega$}\wedge\mbox{\boldmath$\nu$}-
(-1)^{kl}\mbox{\boldmath$\nu$}\wedge\mbox{\boldmath$\omega$}\,)\\
&=&\{\omega_{i_{1}...i_{k}}\; ,\; \nu_{j_{1}...j_{l}}\}\;
dx^{i_{1}}\wedge ...\wedge dx^{i_{k}}\wedge dx^{j_{1}}\wedge
...\wedge dx^{j_{l}}.
\eeas
\bdefin
\mbox{                                                                                  }

\bi
\item \rm A \it connection \rm in $E$ is a linear map 
\mbox{$D:\mbox{Sec}(E)\rightarrow\mbox{Sec}(E^{1})$,} defined by
\be
\forall f\in Sec(E) \h\h\h Df_{\sigma}:=df_{\sigma}+
\frac{1}{i\hbar} A_{\sigma}\ast f_{\sigma}
\ee
where $A_{\sigma}$ is a local 1-form with values in $\Q$ the Lie algebra 
of the group $\eQ$ and $f_{\sigma}:\M\rightarrow \A$ is a representation of the 
section $f$.

\item Any connection $D$ in the bundle $E$ induces connections in bundles $E'$ and
$adj(E)$, respectively (we will use the same symbol D as in each case the connection 
is defined by the same 1-form $A_{\sigma}$)
\beas
\forall f\in Sec(E') \h\h\h Df_{\sigma}:=df_{\sigma}+
f_{\sigma}\ast\frac{1}{i\hbar}A_{\sigma}\\
\forall f\in Sec(adj(E)) \h\h\h Df_{\sigma}:=df_{\sigma}+
\{A_{\sigma}\, ,\,  f_{\sigma}\}
\eeas
\ei
\edefin

If $f_{\sigma}$, $f_{\rho}$ are two representations of the same section
$f:\M\rightarrow E$ then exists $c:\M\rightarrow\eQ$ , such that
$f_{\rho}=c^{-1}\ast f_{\sigma}$. Thus for 
$Df_{\rho}=df_{\rho}+\frac{1}{i\hbar}A_{\rho}\ast f_{\rho}$ one gets 
\be
\label{2.4}
Df_{\rho} = c^{-1}\ast [df_{\sigma}+\frac{1}{i\hbar}A_{\sigma}\ast f_{\sigma}]
\h\h\mbox{where}\h\h A_{\sigma}= c\ast A_{\rho}\ast c^{-1}+
i\hbar\,\,c\ast dc^{-1}.
\ee
The transformation law (\ref{2.4}), can be rewritten for $c^{-1}=e^{\frac{1}{i\hbar}B}$ 
\bdm
A_{\rho}=A_{\sigma}+\sum_{n=1}^{\infty}\frac{1}{n!}
\underbrace{\{B,...,\{B}_{n\mbox{ \tiny times}},A_{\sigma}\}\cdots\}
-dB-\sum_{n=1}^{\infty}\frac{1}{(n+1)!}
\underbrace{\{B,\dots\{B,}_{n\mbox{ \tiny times}}dB\}\cdots\}.
\edm
In what follows we will omit the superscripts $\sigma, \rho$ denoting sections of 
principal bundle. For different representations we will use more common 
symbols \mbox{$'$}, \mbox{$''$}, etc.

\vspace{1 ex}

The connection in $E$ may be extended, in a natural way,  to $E^{k}$. We denote 
the \it exterior covariant differentiation \rm by the same symbol  
$D:\mbox{Sec}(E^{k})\rightarrow\mbox{Sec}(E^{k+1})$
defined for any $\mbox{\boldmath$\omega$}\in\mbox{Sec}(E^{k})$ by \h
\mbox{$D\mbox{\boldmath$\omega$}=d\mbox{\boldmath$\omega$}+
\frac{1}{i\hbar}A\wedge\mbox{\boldmath$\omega$}.$}

The \it curvature \rm of $D$ is a map
$D\circ D\equiv D^{2}:Sec(E)\rightarrow Sec(E^{2})$.
The form $F:=DA$ is a \it curvature form \rm (where operaor 
$D:=d+\frac{1}{i\hbar}A\wedge\h $)
\bdm
F:=dA+\frac{1}{i\hbar}A\wedge A=dA+\frac{1}{2}\{A,A\}.
\edm
For each $\mbox{\boldmath$\omega$}\in\mbox{Sec}(E^{k})$ we have
$D^{2}\mbox{\boldmath$\omega$}=F\wedge\mbox{\boldmath$\omega$}$.
\paragraph{Self-dual Yang-Mills equations.}
The following 2-forms constitute the basis of anti-self-dual forms 
\bdm
\Sigma^{\stackrel{.}{1}\stackrel{.}{1}}:=d\tilde{w}\wedge d\tilde{z},\h
\Sigma^{\stackrel{.}{1}\stackrel{.}{2}}:=\Omega,\h
\Sigma^{\stackrel{.}{2}\stackrel{.}{2}}:=dw\wedge dz
\edm
where $\Omega=g_{\alpha\tilde{\beta}}dz^{\alpha}\wedge dz^{\tilde{\beta}}$ is the K\H{a}hler
form. Then the curvature form $F$ is self-dual iff
\bdm
F\wedge\Sigma^{\stackrel{.}{1}\stackrel{.}{1}}=0\, ,\h
F\wedge\Sigma^{\stackrel{.}{1}\stackrel{.}{2}}=0\, ,\h
F\wedge\Sigma^{\stackrel{.}{2}\stackrel{.}{2}}=0\; .
\edm
The self-dual Yang-Mills (SDYM) equations read
\bea
\label{SDYM 1}
\partial_{w}A_{z}-\partial_{z}A_{w}+\{A_{w}\; ,\; A_{z}\}=0,\\
\label{SDYM 2}
\partial_{\tilde{w}}A_{\tilde{z}}-\partial_{\tilde{z}}A_{\tilde{w}}+
\{A_{\tilde{w}}\; ,\; A_{\tilde{z}}\}=0,\\
\label{SDYM 3}
g^{\tilde{\beta}\alpha}(\partial_{\alpha}A_{\tilde{\beta}}-
\partial_{\tilde{\beta}}A_{\alpha}+
\{A_{\alpha}\; ,\; A_{\tilde{\beta}}\})=0.
\eea
The first two of above equations can be interpreted as integrability conditions 
for substitutions
$A_{\alpha}=i\hbar\; a^{-1}\ast\partial_{\alpha}a$, \h 
$A_{\tilde{\alpha}}=i\hbar\; b^{-1}\ast\partial_{\tilde{\alpha}}b$
where $a,b:{\cal M}\rightarrow \eQ$. Then the third equation
takes the form of
\bdm
i\hbar a^{-1}\ast g^{\tilde{\beta}\alpha}
\partial_{\alpha}[a\ast b^{-1}\ast\partial_{\tilde{\beta}}(b\ast a^{-1})]\ast a=0.
\edm
After the substitiution $J:=b\ast a^{-1}$ this becomes \it  Yang's equation \rm 
(Yang 1977, Parkes 1992)
\be
\label{Yanga}
g^{\tilde{\beta}\alpha}\partial_{\alpha}(J^{-1}\ast\partial_{\tilde{\beta}}J)=0.
\ee
or equivalently $\partial\tilde{\partial}{\cal K}\wedge 
\partial(J^{-1}\ast\tilde{\partial}J)=0$ where 
$\partial=dz^{\alpha}\wedge\partial_{\alpha}$,\h 
$\tilde{\partial}=dz^{\tilde{\alpha}}\wedge\partial_{\tilde{\alpha}}$ are Dolbeault 
operators and ${\cal K}$ is the K\H{a}hler potential i.e. 
$\Omega=\partial\tilde{\partial}{\cal K}$.
This equation arise from a minimum action principle for
$S=\frac{1}{2}\kappa \int\omega^{n}{\cal K}F\wedge F$
where $F$ is a curvatre form of the connection $A=i\hbar J^{-1}\ast\tilde{\partial}J$, 
and $\omega$ is symplectic form on $\Sigma^{2n}$ (Donaldson 1985, Nair and Schiff 1990, 
Mason and Woodhouse 1996).

\vs

In our considerations we will work in the so called K-formalism of
Newman 
(Newman 1978, Leznov 1988, Parkes 1992, Pleba\'{n}ski and Przanowski 1996, Mason and Woodhouse 1996.

Choosing gauge such that 
\mbox{$A=i\hbar J^{-1}\ast\tilde{\partial}J$}, the SDYM equations reduce to Yang's
equation (\ref{Yanga}). It can be rewritten in the form
\be
\label{Parkes}
g^{\tilde{\beta}\alpha}\nabla_{\alpha}A_{\tilde{\beta}}=0
\hspace{1 ex}\Leftrightarrow \hspace{1 ex}
\partial_{\beta}(gg^{\tilde{\alpha}\beta}A_{\tilde{\alpha}})=0
\ee
where $\nabla_{\alpha}$ is the covariant derivative with respect
to the Levi-Civita 
connection\footnote{\label{stopka 1}On a K\H{a}hlerian manifold the hermitian 
connection is at the same time the Levi-Civita connection
\bdm
\Gamma^{\alpha}_{\,\beta\gamma}=
g^{\alpha\tilde{\sigma}}\partial_{\gamma}g_{\beta\tilde{\sigma}}\h ,\h\h
\Gamma^{\tilde{\alpha}}_{\,\tilde{\beta}\tilde{\gamma}}=
g^{\tilde{\alpha}\sigma}\partial_{\tilde{\gamma}}g_{\sigma\tilde{\beta}}.
\edm
This implies that the only non-zero coefficients of  the curvature tensor (apart from 
those obtained from symmetry operations) are
\bdm
R_{\tilde{\alpha}\beta\gamma\tilde{\delta}} =
-g_{\sigma\tilde{\alpha}}\partial_{\tilde{\delta}}\Gamma^{\sigma}_{\beta\gamma}
 = -g_{\beta\tilde{\sigma}}\partial_{\gamma}
\Gamma^{\tilde{\sigma}}_{\tilde{\alpha}\tilde{\delta}}.
\edm
The Ricci tensor \mbox{$R_{\alpha\tilde{\beta}}=g^{\tilde{\sigma}\gamma}
R_{\gamma\tilde{\beta}\alpha\tilde{\sigma}}=
g^{\tilde{\sigma}\gamma}
R_{\alpha\tilde{\beta}\gamma\tilde{\sigma}}=
(\mbox{ln}g),_{\alpha\tilde{\beta}}.$}
and Ricci scalar $R=g^{\tilde{\beta}\alpha}R_{\alpha\tilde{\beta}}$.

The Weyl tensor of conformal curvature 
\bdm
C_{\tilde{\alpha}\beta\gamma\tilde{\delta}}=R_{\tilde{\alpha}\beta\gamma\tilde{\delta}}
+\frac{1}{2}(g_{\beta\tilde{\delta}}R_{\gamma\tilde{\alpha}}+
g_{\gamma\tilde{\alpha}}R_{\beta\tilde{\delta}})-\frac{1}{6}R
g_{\beta\tilde{\delta}}g_{\gamma\tilde{\alpha}}.
\edm
The manifold $({\cal M},ds^{2})$ is called  \it weak heaven \rm (Pleba\'{n}ski 1975) or 
\it right conformally flat \rm (Ko et al. 1981) if the 2-forms \mbox{$C_{\alpha\tilde{\beta}}:=
\frac{1}{2}C_{\alpha\tilde{\beta}\gamma\tilde{\delta}}
dz^{\gamma}\wedge dz^{\tilde{\delta}}$} are selfdual.
For thise to be true  the following conditions should be satisfied: 
\mbox{$C_{\alpha\tilde{\beta}}\wedge\Sigma^{\stackrel{.}{A}\stackrel{.}{B}}=0$,}\h
$\stackrel{.}{A}, \stackrel{.}{B}=\stackrel{.}{1}, \stackrel{.}{2}$.
But the only non-trivial condition is  
$C_{\alpha\tilde{\beta}}\wedge\Sigma^{\stackrel{.}{1}\stackrel{.}{2}}=
\frac{1}{12}Rg_{\alpha\tilde{\beta}}\mbox{\boldmath$\nu$}$\h so $R$ must vanish. 

The space  $({\cal M},ds^{2})$ is \it heavenly \rm if the curvature 2-forms are self-dual.
As $\frac{1}{2}R_{\alpha\tilde{\beta}\gamma\tilde{\delta}}
dz^{\gamma}\wedge dz^{\tilde{\delta}}\wedge\Sigma^{\stackrel{.}{1}\stackrel{.}{2}}=
\frac{1}{2}R_{\alpha\tilde{\beta}}\mbox{\boldmath$\nu$}$
the Ricci tensor must vanish.}
 on ${\cal M}$. Equation (\ref{Parkes}) 
is equivalent to the existence of $\Xi$ such that
\be
\label{def theta}
\partial_{\beta}\Xi=\epsilon_{\beta\gamma}
gg^{\tilde{\alpha}\gamma}A_{\tilde{\alpha}}
\hspace{2 ex}\Rightarrow \hspace{2 ex}
A_{\tilde{\alpha}}=\frac{1}{g}\epsilon^{\beta\gamma}
g_{\gamma\tilde{\alpha}}\partial_{\beta}\Xi=
-\epsilon_{\tilde{\alpha}\tilde{\beta}}g^{\tilde{\beta}\gamma}
\partial_{\gamma}\Xi
\ee
where $\epsilon^{\alpha\beta}$ is a tensor density in $(w,z)$
variables, defined in each coordinate neighbourhood by
$(\epsilon_{\alpha\beta}):=\left(
\begin{array}{cc}
0&1\\
-1&0
\end{array}\right)=:(\epsilon^{\alpha\beta}).$
Analogously $\epsilon_{\tilde{\alpha}\tilde{\beta}}$ is a
tensor density in $(\tilde{w}, \tilde{z})$ variables
$(\epsilon_{\tilde{\alpha}\tilde{\beta}}):=\left(
\begin{array}{cc}
0&1\\
-1&0
\end{array}\right)=:
(\epsilon^{\tilde{\alpha}\tilde{\beta}}).$
The definition (\ref{def theta}) of $\Xi$ implies that under the  change of 
variables $\tilde{w}'=\tilde{w}'(\tilde{w},\tilde{z})$, 
$\tilde{z}'=\tilde{z}'(\tilde{w},\tilde{z})\,$ \hspace{.5 ex} 
$\Xi$ transforms according to
$\Xi'=\frac{\partial (\tilde{w},\tilde{z})}
{\partial (\tilde{w'},\tilde{z'})}\, \Xi $
i.e. $\Xi$ is in these variables a scalar density. In this case covariant 
derivative $\nabla_{\tilde{\alpha}}$ acts on densities according to the rule
\mbox{$\nabla_{\tilde{\alpha}}\Xi=\partial_{\tilde{\alpha}}\Xi-
(\ln g),_{\tilde{\alpha}}\Xi$}\h while
\mbox{$\nabla_{\alpha}\Xi=\partial_{\alpha}\Xi $} 

Inserting $A_{\tilde{\alpha}}$ given by (\ref{def theta})
into (\ref{SDYM 2}) one gets 
\be
\label{Master}
g^{\tilde{\alpha}\beta}\nabla_{\tilde{\alpha}}\nabla_{\beta}\Xi
+\frac{1}{2g}\epsilon^{\alpha\beta}
\{\nabla_{\alpha}\Xi\, ,\,\nabla_{\beta}\Xi\}=0
\ee
For the first time this equation was proposed by Q-H. Park (1992) for Poisson algebra.  
In the case of Moyal agebra and flat base manifold it was considered by 
Pleba\'{n}ski and Przanowski (1996), Przanowski and Forma\'{n}ski (1999). 

The linearized equation (\ref{Master}) reads
\be
\label{Linear Master}
g^{\tilde{\alpha}\beta}\nabla_{\tilde{\alpha}}\nabla_{\beta}\eta
+\frac{1}{g}\epsilon^{\alpha\beta}
\{\nabla_{\alpha}\Xi\, ,\,\nabla_{\beta}\eta\}=0
\ee

Let us consider a special case of the base manifold $\M$ i.e. the heavenly space.
For a K\H{a}hlerian manifold to be heavenly space it is necessary and sufficient 
that the Ricci tensor $R_{\alpha\tilde{\beta}}$ 
vanish\footnote{compare footnote \ref{stopka 1}}  i.e. $(\ln g),_{\alpha\tilde{\beta}}=0$. 
Thus the determinant 
$g=G(w,z)\tilde{G}(\tilde{w},\tilde{z})$, and rewriting (\ref{def theta})
once again 
\bdm
A_{\tilde{\alpha}}=\frac{1}{G}\epsilon^{\beta\gamma}
g_{\gamma\tilde{\alpha}}\partial_{\beta}(\frac{\Xi}{\tilde{G}})
=-\tilde{G}\epsilon_{\tilde{\alpha}\tilde{\beta}}
g^{\tilde{\beta}\gamma}\partial_{\gamma}(\frac{\Xi}{\tilde{G}})
\edm
we can define $\Theta:=\frac{\Xi}{\tilde{G}}$ which is now a 
scalar function. With help of this function the master equation 
(\ref{Master}) reads now
\be
\label{ME in H}
g^{\tilde{\beta}\alpha}\partial_{\tilde{\beta}}\partial_{\alpha}\Theta
+\frac{1}{2G}\epsilon^{\alpha\beta}\{\partial_{\alpha}\Theta\, ,\,
\partial_{\beta}\Theta\}=0
\ee
It will be called \it master equation \rm (ME). This equation arises from a minimum 
action principle for the action
\be
\label{action for ME}
S=\frac{1}{2}\int\omega^{n}[\tilde{G}\tilde{\epsilon}\wedge\frac{1}{3}
\Theta\ast\{\partial\Theta ,\partial\Theta\}+
\Omega\wedge\partial\Theta\wedge\tilde{\partial}\Theta]
\ee
where $\tilde{\epsilon}:=\frac{1}{2}\epsilon_{\tilde{\alpha}\tilde{\beta}}
dx^{\tilde{\alpha}}\wedge dx^{\tilde{\beta}}$ and 
$\Omega=g_{\alpha\tilde{\beta}}dz^{\alpha}\wedge dz^{\tilde{\beta}}$
is the K\H{a}hler form.

\noindent (Note that our notation is covariant and we do not work in 
any special coordinates, like for example Pleba\'{n}ski`s coordinates
(Pleba\'{n}ski 1975) in which g=1 (Parkes 1992).)

\noindent The action (\ref{action for ME}) generalizes actions given 
by Boyer and Pleba\'{n}ski (1985), Leznov (1988), Parkes (1992), 
Pleba\'{n}ski and Przanowski (1996). 

\section{Conservation laws and twistor construction.}
\setcounter{equation}{0}
In this section we construct two hierarchies of conservation laws for ME
on heavenly background (\ref{ME in H}). It is known  from the previous section 
that in the algebra $\A$ there exist both the $\ast$-product and the 
bracket $\{\cdot, \cdot\}$. Hence one can expect the existence of two different 
linear systems for ME.
\paragraph{Hierarchy of hidden symmetries of ME.} 
Let ${\cal W}$ denote the space of solutions to the master equation (ME) (\ref{ME in H}), 
and $T{\cal W}$ the space of solutions to the \it linearized master equation \rm (LME)
\be
\label{LME in H}
g^{\tilde{\beta}\alpha}\partial_{\tilde{\beta}}\partial_{\alpha}\phi
+\frac{1}{G}\epsilon^{\alpha\beta}\{\partial_{\alpha}\Theta\, ,\,
\partial_{\beta}\phi\}=0.
\ee
Define two operators acting on functions on $\M$ with values in $\A$
\be
{\cal L}^{\alpha}:=g^{\tilde{\beta}\alpha}\partial_{\tilde{\beta}}
+\frac{\epsilon^{\beta\alpha}}{G}\{\partial_{\beta}\Theta ,\,\cdot\,\}\, ,
\h\h\alpha=1, 2.
\ee
Then their commutator can be easily found to be
\bdm
[{\cal L}^{w},{\cal L}^{z}](\cdot)=
\frac{1}{G}\{
g^{\tilde{\beta}\alpha}\partial_{\tilde{\beta}}\partial_{\alpha}\Theta
+\frac{1}{2G}\epsilon^{\alpha\beta}\{\partial_{\alpha}\Theta\, ,\,
\partial_{\beta}\Theta\}\, ,\,\cdot\,\}.
\edm
So the operators ${\cal L}^{\alpha}$ commute iff $\Theta$ satisfies ME (\ref{ME in H}).
Eq. (\ref{LME in H}) written in terms of these operators reads 
\bdm
{\cal L}^{\alpha}\partial_{\alpha}\phi=0
\edm
Suppose that $\phi_{(0)}\in T{\cal W}$. Then define the \it current \rm 
$J_{(1)}$ with components $J^{\alpha}_{(1)}:={\cal L}^{\alpha}\phi_{(0)}$, 
$J^{\tilde{\alpha}}_{(1)}=0$. Thus 
\mbox{$\nabla_{i}J^{i}_{(1)}=
\nabla_{\alpha}J^{\alpha}_{(1)}=\nabla_{\alpha}{\cal L}^{\alpha}\phi_{(0)}
={\cal L}^{\alpha}\partial_{\alpha}\phi_{(0)}$}.
As $\phi_{(0)}$ solves the LME 
the current $J_{(1)}$ fulfills the conservation law 
$\nabla_{\alpha}J^{\alpha}_{(1)}=0$. This conservation law
can be written in the form 
$\partial_{\alpha}(GJ^{\alpha}_{(1)})=0$, which implies existence 
of the scalar function $\phi_{(1)}$, such that
\be
\label{defin of phi1}
GJ^{\alpha}_{(1)}=-\epsilon^{\alpha\beta}\partial_{\beta}\phi_{(1)}
\hspace{1 ex}\Rightarrow\hspace{1 ex}
\partial_{\alpha}\phi_{(1)}=G\epsilon_{\alpha\beta}J^{\beta}_{(1)}
\ee
This function gives rise to the next current 
$J^{\alpha}_{(2)}:={\cal L}^{\alpha}\phi_{(1)}$, which divergence
also vanishes, i.e.
\beas
\nabla_{\alpha}J^{\alpha}_{(2)}&=&
\nabla_{\alpha}{\cal L}^{\alpha}\phi_{(1)}=
{\cal L}^{\alpha}\partial_{\alpha}\phi_{(1)}=
\epsilon_{\alpha\beta}{\cal L}^{\alpha}G{\cal L}^{\beta}\phi_{(0)}\\
&=&\{g^{\tilde{\beta}\alpha}\partial_{\tilde{\beta}}\partial_{\alpha}\Theta
+\frac{1}{2G}\epsilon^{\alpha\beta}\{\partial_{\alpha}\Theta\, ,\,
\partial_{\beta}\Theta\}\, ,\, \phi_{(0)}\}
\stackrel{\mbox{\footnotesize{by ME}}}{=}0
\eeas
Note that the above equality also says that $\phi_{(1)}\in T{\cal W}$. One can 
repeat above construction starting from $\phi_{(1)}$. We are led to an iterative 
procedure. Given the n-th conserved charge $\phi_{(n)}$ one constructs the (n+1) current 
$J^{\alpha}_{(n+1)}:={\cal L}^{\alpha}\phi_{(n)}$ and then one solves
$\partial_{\alpha}\phi_{(n+1)}=G\epsilon_{\alpha\beta}J^{\beta}_{(n+1)}$
for $\phi_{(n+1)}$ i.e. $\phi_{(n+1)}$ is a solution  of 
\be
\label{defin of R}
\partial_{\alpha}\phi_{(n+1)}=G\epsilon_{\alpha\beta}
{\cal L}^{\beta}\phi_{(n)}.
\ee
Such a solution is an element of $T{\cal W}$ and it defines a divergence free current. 

\noindent\bf Remarks. \rm
\begin{itemize}
\item In this way we define an integro-differential \it recursion \rm operator  
$\phi_{(n+1)}={\cal R}\phi_{(n)}$ depending additionaly on boundary conditions 
imposed. This operator is invertible.
\item The elements of $T{\cal W}$ are hidden symmetries of ME. This  symmetries 
generalize the symmetries for heavenly equations obtained by 
Boyer and Pleba\'{n}ski (1985), Strachan (1993), Husain (1994), Dunajski and Mason (2000).
\end{itemize}
\paragraph{Second collection of conserved charges.}
The existence in each fibre of the bundle $E$ of the $\ast$-product
allows us to construct another set of operators
\bdm
{\cal D}^{\alpha}:=g^{\tilde{\beta}\alpha}\partial_{\tilde{\beta}}
+\frac{1}{i\hbar}\frac{\epsilon^{\beta\alpha}}{G}
\partial_{\beta}\Theta\ast
\edm
where $\Theta$ is a solution of ME (\ref{ME in H}).
For any function with value in $\A$ we have 
\mbox{$\nabla_{\alpha}{\cal D}^{\alpha}-{\cal D}^{\alpha}\partial_{\alpha}=0$.}
Let $\eta_{(0)}$ be a solution to 
\be
\label{drugie liniowe}
{\cal D}^{\alpha}\partial_{\alpha}\eta_{(0)}=0.
\ee
Then define vector $j^{\alpha}_{(1)}:={\cal D}^{\alpha}\eta_{(0)}$,
$j^{\tilde{\alpha}}:=0$. The  divergence of $j^{i}_{(1)}$ vanishes
\bdm
\nabla_{i}j^{i}_{(1)}=\nabla_{\alpha}j^{\alpha}_{(1)}=
\nabla_{\alpha}{\cal D}^{\alpha}\eta_{(0)}=
{\cal D}^{\alpha}\partial_{\alpha}\eta_{(0)}=0.
\edm
In the same way as in the previous case the conserved current $j^{i}_{(1)}$ 
defines a function $\eta_{(1)}$ by a system of equations
\be
\partial_{\alpha}\eta_{(1)}=G\epsilon_{\alpha\beta}j^{\beta}_{(1)}.
\ee
This function fulfills 
\beas
{\cal D}^{\alpha}\partial_{\alpha}\eta_{(1)} & = &
{\cal D}^{\alpha}G\epsilon_{\alpha\beta}
j^{\beta}_{(1)}=\epsilon_{\alpha\beta}{\cal D}^{\alpha}
G{\cal D}^{\beta}\eta_{(0)}\\
& = & \frac{1}{i\hbar}(
g^{\tilde{\beta}\alpha}\partial_{\tilde{\beta}}\partial_{\alpha}\Theta
+ \frac{\epsilon^{\alpha\beta}}{2G}\{\partial_{\alpha}\Theta,
\partial_{\beta}\Theta\})\ast\eta_{(0)}
\stackrel{\mbox{\footnotesize{by ME}}}{=}0.
\eeas
The function $\eta_{(1)}$ fulfills the same equation as $\eta_{(0)}$.
This allows us to define another current  $j^{i}_{(2)}$, 
with components $j^{\alpha}_{(2)}={\cal D}^{\alpha}\eta_{(1)}$,
$j^{\tilde{\alpha}}_{(2)}=0$. The divergence of $j_{(2)}$ vanishes.

Continuing this procedure we arrive at the series of conserved
charges $\eta_{(0)},\eta_{(1)},\dots $ and  currents
$j^{\alpha}_{(1)},j^{\alpha}_{(2)},\dots $ defined by the recursion 
equations
\be
\label{recursion 2}
\partial_{\alpha}\eta_{(n+1)}=G\epsilon_{\alpha\beta}j^{\beta}_{(n+1)}=
G\epsilon_{\alpha\beta}{\cal D}^{\beta}\eta_{(n)}\, ,
\hspace{1.5 ex}n=0,1,\dots .
\ee
\bf Remarks\rm 
\bi
\item
As in the case of hidden symmetries the system (\ref{recursion 2}) defines 
the recursion operator $\eta_{(n+1)}=\widetilde{{\cal R}}\eta_{(n)}$.
The above hierarchy of conservation laws is characteristic for self-dual Yang-Mills 
equations (compare (Brezin et al. 1979, Prasad et al. 1979, Chau 1983)).
\item 
Both hierarchies were presented in Przanowski et al. 2001(a), Przanowski et al. 2001(b).
in the case of the complexified Minkowski space $\M$ and the Moyal $\ast$-product. 
\ei

The characteristic feature of integrable systems besides the existence of 
infinite number of conservation laws is the existence of a Lax pair and 
some geometric construction related to the system considered. We are going to deal 
with this problem.
\paragraph{Twistors for $\M$.}
\it Twistor surface \rm or \it $\beta$-plane \rm or \it null string \rm (Penrose 1976, 
Pleba\'{n}ski and Hacyan 1975, Flaherty 1976, Ward and Wells 1990, Mason and Woodhouse 1996)
is a 2-dimmensional submanifold \mbox{${\cal S}\subset{\cal M}$} such that

$\bullet$ ${\cal S}$ is totally null i.e. $\forall p\in{\cal S}$\h and
$\forall v\in T_{p}{\cal S}$ $ds^{2}(v,v)=0$,

$\bullet$ The 2-form orthogonal
to ${\cal S}$ is anti-self-dual.

\noindent This implies that it is also totally geodesic i.e. $\forall p\in{\cal S}$\h 
and $\forall v\in T_{p}{\cal S}$ geodesic with tangent vector $v$ in $p$
lies on the surface ${\cal S}$.

\vspace{1 ex}

For heavenly space $({\cal M}, ds^{2})$ we have 
$\partial_{\alpha}\partial_{\tilde{\beta}}\ln g=0$ i.e. $g=G(w,z)\tilde{G}(\tilde{w},\tilde{z})$.
In appropriate coordinates $g=1$ and this is the \it first heavenly equation  \rm (Pleba\'{n}ski 1975).
We work in an arbitrary coordinate system, which means that the determinant $g$ is a product
of two functions.

For each \mbox{$\lambda\in\mbox{\boldmath$CP$}^{1}-\{\infty\}$} 
the integral 2-surface of two vetor fields
\bea
\ell_{w} & = & \frac{\partial}{\partial w} -
\lambda G g^{\tilde{\sigma}z}\frac{\partial}{\partial z^{\tilde{\sigma}}},\nonumber\\
\label{wektory ell}
\ell_{z} & = & \frac{\partial}{\partial z} +
\lambda G g^{\tilde{\sigma}w}\frac{\partial}{\partial z^{\tilde{\sigma}}}.
\eea
is a twistor surface. This follows from the Frobenius theorem as those fields commute, and 
from the fact that 
$ds^{2}(\ell_{w},\ell_{w})=
ds^{2}(\ell_{w},\ell_{z})=
ds^{2}(\ell_{z},\ell_{z})=0$.
The anti-self-dual form  \h
$\Sigma(\lambda):=\tilde{G}d\tilde{w}\wedge d\tilde{z}-\lambda\Omega+
\lambda^{2}Gdw\wedge dz$ \h is orthogonal to 
the distribution ${\cal W}_{\lambda}=\mbox{span}\{\ell_{w}, \ell_{z}\}$. Moreover 
it is closed $d\Sigma(\lambda)=0$ and degenerate $\Sigma(\lambda)\wedge\Sigma(\lambda)=0$
From Daurboux theorems this allows one to introduce smooth functions $P^{w}, P^{z}$
such that $\Sigma(\lambda)$ takes the canonocal form $\Sigma(\lambda)=dP^{w}\wedge dP^{z}$. 

Analogously in the domain $\mbox{\boldmath$CP$}^{1}-\{0\}\ni\zeta$, the twistor surface  
is defined by
\bea
\underline{\ell_{w}} & = & \frac{1}{\zeta}\frac{\partial}{\partial w} -
G g^{\tilde{\sigma}z}\frac{\partial}{\partial z^{\tilde{\sigma}}}\nonumber\\
\underline{\ell_{z}} & = & \frac{1}{\zeta}\frac{\partial}{\partial z}+
G g^{\tilde{\sigma}w}\frac{\partial}{\partial z^{\tilde{\sigma}}}.
\eea
and \h $\Sigma(\zeta)=\frac{1}{\zeta^{2}}\tilde{G}d\tilde{x}\wedge d\tilde{y}-
\frac{1}{\zeta}\Omega + G dx\wedge dy$. \h  The canonical form of $\Sigma(\zeta)$ reads
\mbox{$\Sigma(\zeta)=d\underline{P^{w}}\wedge d\underline{P^{z}}$.}

On the patching $\mbox{\boldmath$CP$}^{1}-\{0,\infty\}$ for $\lambda=\zeta$ 
the distributions considered are equivalent. Then we see that \it for each point $p\in\M$ 
and for each $\lambda\in\mbox{\boldmath$CP$}^{1}$ there exists a twistor surface through $p$ \rm  
(Penrose 1976). The set $\PT$ of all twistor surfaces is a $3$-dimmensional complex
manifold called the \it projective twistor space\rm . It is covered by two coordinate 
neighbourhoods $(V,(P^{w}, P^{z},\lambda))$ for 
$\lambda\in\mbox{\boldmath$CP$}^{1}-\{\infty\}$ 
and $(\underline{V},(\underline{P^{w}}, \underline{P^{z}},\zeta))$ for 
$\zeta\in\mbox{\boldmath$CP$}^{1}-\{0\}$.

More general result also holds i.e. the projective twistor space exists iff
$({\cal M}, ds^{2})$ is a weak heaven (Penrose and Ward 1980).

Both manifolds $\M$ and $\PT$ are embedded in the so called \it correspondence space \rm
${\cal F}:=\M\times\mbox{\boldmath$CP$}^{1}$ 

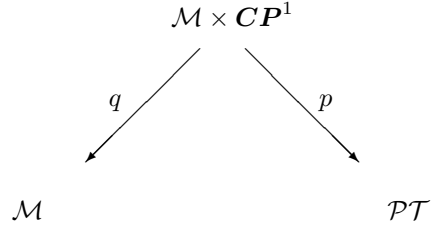
\setlength{\unitlength}{5ex}
\begin{picture}(9,5)
\put(2.9,3){$q$}
\put(4,4.4){$\M\times\mbox{\boldmath$CP$}^{1}$}
\put(4.5,4){\vector(-1,-1){2}}
\put(1.2,1){$\M$}
\put(7.8,1){$\PT$}
\put(6.6,3){$p$}
\put(5.3,4){\vector(1,-1){2}}
\end{picture}

\paragraph{The Lax pair and Penrose-Ward transform.}
In this paragraph we construct the formal bundle over twistor space $\PT$ which is determined 
by a solution $\Theta$ of master equation (ME). 
First we start with a Lax pair for ME. For each value of a spectral parameter 
belonging to $\mbox{\boldmath$CP$}^{1}$, consider a pair of operators
\beas
M_{w}&=&\partial_{w}-\lambda Gg^{\tilde{\sigma}z}\partial_{\tilde{\sigma}}
-\frac{\lambda}{i\hbar}\partial_{w}\Theta\ast =\,
\ell_{w}-\frac{\lambda}{i\hbar}\partial_{w}\Theta\ast\\
M_{z}&=&\partial_{z}+\lambda Gg^{\tilde{\sigma}w}\partial_{\tilde{\sigma}}
-\frac{\lambda}{i\hbar}\partial_{z}\Theta\ast =\,
\ell_{z}-\frac{\lambda}{i\hbar}\partial_{z}\Theta\ast\h ,\h\h\h
\mbox{for}\h\h\h\lambda\in\mbox{\boldmath$CP$}^{1}-\{\infty\}.
\eeas
And respectively
\beas
\underline{M_{w}}&=&\frac{1}{\zeta}\partial_{w}- Gg^{\tilde{\sigma}z}\partial_{\tilde{\sigma}}
-\frac{1}{i\hbar}\partial_{w}\Theta\ast =\,
\underline{\ell_{w}}-\frac{1}{i\hbar}\partial_{w}\Theta\ast\\
\underline{M_{z}}&=&\frac{1}{\zeta}\partial_{z}+Gg^{\tilde{\sigma}w}\partial_{\tilde{\sigma}}
-\frac{1}{i\hbar}\partial_{z}\Theta\ast =\,
\underline{\ell_{z}}-\frac{1}{i\hbar}\partial_{z}\Theta\ast\h ,\h\h\h\mbox{for}\h\h\h\zeta\in
\mbox{\boldmath$CP$}^{1}-\{0\}.
\eeas
Then one has
\bdm
\epsilon^{\alpha\beta}M_{\alpha}\,M_{\beta}=\frac{\lambda^{2}}{i\hbar}
\, G\,[g^{\tilde{\beta}\alpha}\partial_{\tilde{\beta}}\partial_{\alpha}\Theta+
\frac{1}{i\hbar\, G}\epsilon^{\alpha\beta}\partial_{\alpha}\Theta\ast\partial_{\beta}\Theta\, ].
\edm
Thus for each $\lambda\in\mbox{\boldmath$CP$}^{1}-\{\infty\}$ this commutator vanishes iff  
$\Theta$ satisfies ME.
Analogously for $\zeta\in\mbox{\boldmath$CP$}^{1}-\{0\}$ 
\mbox{$[\underline{M_{w}},\,\underline{M_{z}}]=0$.} iff $\Theta$ satisfies master equation. 

If $\Theta$ is any solution of ME then Frobenius integrability conditions are satisfied 
and one can find a solution of the linear system
\be
\label{Laxa}
M_{w}\,\Psi(\lambda)=0 \h , \h\h\h
M_{z}\,\Psi(\lambda)=0.
\ee
where \mbox{$\Psi(\lambda)\equiv\Psi(t,\hbar;w,z,\tilde{w},\tilde{z},\lambda)\in\A$}.
In particular this solution is analytic in $\lambda$ in some naighbourhood of 
$0\in\mbox{\boldmath$CP$}^{1}$. We will construct such a solution from conserved charges.

Let $\eta_{(k)}$\h $k=0,1,2, ...$ denote conserved charges defined by recursion 
relations (\ref{recursion 2}). For $\lambda\in\mbox{\boldmath$CP$}^{1}-\{\infty\}$
we define 
\be
\Psi(\lambda):=\sum_{k=0}^{\infty}\lambda^{k}\eta_{(k)}(t,\hbar;w,z,\tilde{w},\tilde{z}).
\ee
The conserved charges can be chosen such that  the radius of convergence is grater 
then zero. As all $\eta_{(k)}$ satisfies  (\ref{recursion 2}) thus above $\Psi(\lambda)$ 
satisfies (\ref{Laxa}).

\vspace{1 ex}

By a \it fundamental solution \rm of the system (\ref{Laxa}) we mean a solution with 
value in the group $\eQ$. Taking appropriate solution $\eta_{(0)}$ of 
(\ref{drugie liniowe}) we get  $\Psi(\lambda)$  with free element equal to $1$. In particular 
if we take $\eta_{(0)}=1$ then recursion relation give 
$\eta_{(1)}=\frac{1}{i\hbar}\Theta$.

\vspace{1 ex}

Two fundamental solutions $\Psi_{1}(\lambda)$ and $\Psi_{2}(\lambda)$ differ only by a 
twistor function i.e. $\Psi_{1}(\lambda)=\Psi_{2}(\lambda)\ast {\cal H}$ where 
${\cal H}:{\cal F}\rightarrow\eQ$ is constant along each twistor surface  
$\ell_{\alpha}{\cal H}=0$, $\alpha=w,z$. 

\vspace{1 ex}

For $\zeta\in\mbox{\boldmath$CP$}^{1}-\{0\}$, using another sequence of conserved charges 
$\{\eta'_{(k)}\}_{k=0}^{\infty}$,  one can construct a fundamental solution of the system
\bea
\label{Laxa2}
\underline{M_{w}}\,\underline{\Psi}(\zeta)=0 \h\h\h & , &
\underline{M_{z}}\,\underline{\Psi}(\zeta)=0\nonumber\\
\underline{\Psi}(\zeta) &=&\eta'_{(0)}+\sum_{k=1}^{\infty}(\frac{1}{\zeta})^{k}\eta'_{(k)}.
\eea
This time  $\{\eta'_{(k)}\}_{k=0}^{\infty}$ is a sequence for which 
$\eta'_{(k)}=\widetilde{{\cal R}}\eta'_{(k+1)}$, $k\geq 1$.  The operator 
$\widetilde{\cal R}$ is defined by (\ref{recursion 2}). Moreover $\eta'_{(0)}$ fulfills 
additionally ${\cal D}^{\alpha}\eta'_{(0)}=0$ which are the same as those for $J^{-1}$ in 
Yang's equation (\ref{Yanga}) (compare (Mason and Woodhouse 1996)).

\vspace{1 ex}

On the overlap of domains, for $\lambda=\zeta\in\mbox{\boldmath$CP$}^{1}-\{0,\infty\}$, 
$\Psi(\lambda)=\underline{\Psi}(\lambda)\ast{\cal H}$ where ${\cal H}$ is a twistor 
function defined uniquely by $\Psi(\lambda)$ and $\underline{\Psi}(\lambda)$.
As it takes values in the group $\eQ$ and twistor space may be covered only by those two 
neighbourhood the knowledge of this function is sufficient to recover a bundle over $\PT$
with ${\cal H}$ as a transition function. In this way each solution of (ME) corresponds to
one bundle over the space $\PT$.
\paragraph{Dressing operator.}
The hidden symmetries i.e elements of the space $T{\cal W}$ of solutions to Eq. 
(\ref{LME in H}) define a second Lax pair 
\be
\label{tilde phi}
\Phi(\lambda):=\sum_{n=0}^{\infty}\lambda^{n}\phi_{(n)}\h ,\h\h
\lambda\in\mbox{\boldmath$CP$}^{1}-\{\infty\}
\ee
As all $\phi_{(n)}$, $n=0,1,...$ satisfy LME (\ref{LME in H}) 
$\Phi(\lambda)$ satisfies the system
\be
\label{Lax-symetrii 1}
\partial_{\alpha}\Phi(\lambda)-\lambda G\epsilon_{\alpha\beta}
{\cal L}^{\beta}\Phi(\lambda)=0\h ,\h\h
\lambda\in\mbox{\boldmath$CP$}^{1}-\{\infty\},
\ee
which is a Lax pair for ME.

Respectively in the neighbourhood of infinity $\zeta\in\mbox{\boldmath$CP$}^{1}-\{0\}$ 
\be
\label{underline phi}
\underline{\Phi}(\zeta)=
\sum_{n=0}^{\infty}(\frac{1}{\zeta})^{n}\phi_{(n)}, \h\h
\zeta\in\mbox{\boldmath$CP$}^{1}-\{0\}
\ee
then the spectral system tis of the form
\be
\label{Lax-symetrii 2}
\frac{1}{\zeta}\partial_{\alpha}\underline{\Phi}(\zeta) -
G\epsilon_{\alpha\beta}{\cal L}^{\beta}
\underline{\Phi}(\zeta)=0.
\ee
Let $F(\lambda):=F(t,\hbar;w,z,\tilde{w},\tilde{z},\lambda)$ be such that
\be
\label{definitionofF}
\Phi(\lambda)=\Psi(\lambda)\ast F(\lambda)\ast
\Psi^{-1}(\lambda).
\ee
Such $F(\lambda)$  exists and is uniquely defined by 
$\Psi(\lambda)$ and $\Phi(\lambda)$. Moreover, as 
$\Psi(\lambda)$ and  $\Phi(\lambda)$ fulfill (\ref{Laxa}) and
(\ref{Lax-symetrii 1}) respectively, $F(\lambda)$ 
has to be constant along each twistor surface i.e. it depends only on
$(P^{w},\,P^{z},\lambda)$

The definition (\ref{definitionofF}) describing $F(\lambda)$ constitutes
$\Psi(\lambda)$ as a \it dressing operator \rm for a linear system
(\ref{Lax-symetrii 1}).
\paragraph{Algebra of hidden symmetries.}
Consider a superposition of solutions to the linearized master equation (\ref{LME in H}),
written in terms of the above defined dressing operator 
\bea
\label{gensymmofsolutME}
\delta_{(F\underline{F})}\Theta & = & \frac{1}{2\pi i}
\oint_{\gamma}\frac{d\lambda}{\lambda^{2}}
(-\Psi(\lambda)\ast F(\lambda)\ast\Psi^{-1}(\lambda)\, + \,
\underline{\Psi}(\lambda)\ast \underline{F}(\lambda)\ast
\underline{\Psi}^{-1}(\lambda))\nonumber\\
\mbox{where}
 &  & F(\lambda ) =
F(t,\hbar; \widetilde{P^{w}},\widetilde{P^{z}},\lambda),  \nonumber \\
 &  & \underline{F}(\lambda) =
\underline{F}(t,\hbar; \underline{P^{w}},\underline{P^{z}},\lambda).
\eea
(Compare (Park 1990 and 1992, Takasaki 1990).)
The contour $\gamma$ in (\ref{gensymmofsolutME}) is the boundary of a domain 
containing $\lambda=0$ and it does not cross any singularity
of integrated functions.

To find an algebra of hidden symmetries consider a \it commutator \rm
\bea
\lefteqn{[\, \delta_{(F_{1}\underline{F_{1}})}\, ,\,
\delta_{(F_{2}\underline{F_{2}})} ] \Theta =}\\
& & \delta_{(F_{1}\underline{F_{1}})}(\Theta +
\delta_{(F_{2}\underline{F_{2}})}\Theta)\; -\;
\delta_{(F_{1}\underline{F_{1}})}\Theta\;-\;
\delta_{(F_{2}\underline{F_{2}})}(\Theta+
\delta_{(F_{1}\underline{F_{1}})}\Theta)\; +\;
\delta_{(F_{2}\underline{F_{2}})}\Theta \nonumber
\eea
The following theorem holds (compare (Takasaki 1990, Park 1992,
Dunajski and Mason 2000))

The hidden symmetries of ME constitute the algebra 
\be
\label{algebraofhiddensymmetries}
[\, \delta_{(\widetilde{F_{1}}\underline{F_{1}})}\, ,\,
\delta_{(\widetilde{F_{2}}\underline{F_{2}})}]\Theta\, =\,
\delta_{(\{\widetilde{F_{1}},\widetilde{F_{2}}\}
\{\underline{F_{1}},\underline{F_{2}}\})}\Theta
\ee
The proof can be found in (Przanowski et al. 2001(b), Forma\'{n}ski 2004)

\setcounter{equation}{0}
\section{Integrability of ME.}
\paragraph{The homogenous Hilbert problem for formal power series.} 
\label{szeregi Hilbert}
The Hilbert problem for formal power series can be defined in the similar form
as in the case of vector functions. The latter case can be found in the monographs 
(Muscheliszwili 1962, Pogorzelski 1966). We will show the existence theorem in the
first case.

Let $L$ be a smooth contour. Let $S^{+}$ denote the interior of $L$ and let $0\in S^{+}$. By 
$S^{-}$ we denote the exterior of L i.e. $S^{-}:=\mbox{\boldmath$CP$}^{1}-S^{+}-L$. 

Let the formal power series 
\bdm
\Phi(t,\hbar;\lambda)=\sum_{m=0}^{\infty}\sum_{k=-m}^{\infty}t^{m}\hbar^{k}\Phi_{m,k}(\lambda)
\edm
be such that all the functions $\Phi_{m,k}(\lambda)$ are \it sectionally holomorphic \rm 
what means that each $\Phi_{m,k}(\lambda)$ is holomorphic on  $S^{+}$ and $S^{-}$.
The formal power series is said to have a \it finite degree at infinity \rm if for each 
function $\Phi_{m,k}(\lambda)$ there exist $c_{m,k}\in\mbox{\boldmath$Z$}$ such that 
$\lim_{|\lambda|\rightarrow\infty}\frac{|\Phi_{m,k}(\lambda)|}{|\lambda|^{c_{m,k}}}=0$. In case
$c_{m,k}>0$ in the neighbourhood of infinity we can write
\bdm
\Phi_{m,k}(\lambda)=\gamma_{m,k}(\lambda)+O(\frac{1}{\lambda})\h\h\mbox{where}\h\h
\gamma_{m,k}(\lambda)\mbox{ is a polynomial}
\edm
For $c_{m,k}<0$ we have $\gamma_{m,k}(\lambda)=0$ and for $c_{m,k}=0$ the $\gamma$'s
are constant. The formal power series 
$\gamma(t,\hbar;\lambda)=
\sum_{m=0}^{\infty}\sum_{k=-m}^{\infty}t^{m}\hbar^{k}\gamma_{m,k}(\lambda)$
will be called the \it principal part at infinity \rm of the series $\Phi(t,\hbar;\lambda)$.
It is said that the series $\Phi(t,\hbar;\tau)$ $\tau\in L$ satisfies on L the \it H\H{o}lder 
condition \rm $H(\alpha)$, $0<\alpha\leq 1$ if there exist constants
$A_{m,k}$ such that $\forall \tau_{1},\tau_{2}\in L$ \h 
\mbox{$|\Phi_{m,k}(\tau_{2})-\Phi_{m,k}(\tau_{1})|\leq A_{m,k}|\tau_{2})-\tau_{1}|^{\alpha}$.}

The homogeneous Hilbert problem can be formulated as follows.
Suppose we are given an element $G(t,\hbar;\xi)$ of the group $\eQ$, defined on $L$ and 
satisfying the H\H{o}lder condition on $L$. 
Find  a sectionally holomorphic  formal power 
series $\Phi(t,\hbar;\lambda)$ having finite degree at infinity, continous on $L$ and 
satisfying the boundary condition
\be
\label{zagad pierwsze}
\Phi^{+}(t,\hbar;\xi)=\Phi^{-}(t,\hbar;\xi)\ast G(t,\hbar;\xi)\h\h\h\h\xi\in L
\ee
$\Phi^{+}(t,\hbar;\xi)$ and $\Phi^{-}(t,\hbar;\xi)$ denote the limit values i.e.
\beas
\Phi^{+}(t,\hbar;\xi)=\lim_{\lambda\rightarrow\xi}\Phi(t,\hbar;\lambda)\h\h\mbox{for }
\lambda\in S^{+}\\
\Phi^{-}(t,\hbar;\xi)=\lim_{\lambda\rightarrow\xi}\Phi(t,\hbar;\lambda)\h\h\mbox{for }
\lambda\in S^{-}
\eeas
We will seek for  a solution of this problem in the class of formal power series 
satisfying H\H{o}lder condition on $L$.

In the case of finite groups this problem is solved by the \it Birkhoff factorization
theorem \rm (Birkhoff 1913, Mason and Woodhouse 1996). 

Since $\Phi^{+}(t,\hbar;\xi)$ is the limit value of $\Phi(t,\hbar;\lambda)$ holomorphic in 
$S^{+}$, from the Cauchy theorem we find
\bdm
0=\frac{1}{2\pi i}\int_{L}\frac{\Phi^{+}(t,\hbar;\tau)}{\tau-\lambda}d\tau\h\h\h\lambda\in S^{-}, 
\edm
By the Plemelj formula (Plemelj 1908), in the limit $\lambda\rightarrow\xi\in L$
one gets
\be
\label{C1}
-\frac{1}{2}\Phi^{+}(t,\hbar;\xi)+
\frac{1}{2\pi i}\int_{L}\frac{\Phi^{+}(t,\hbar;\tau)}{\tau-\xi}d\tau=0
\ee
where the integral above is taken in the sense of principal value. The equation (\ref{C1}) is an 
integral equation for $\Phi^{+}(t,\hbar;\xi)$.

Analogously the Cauchy theorem guarantees that $\Phi^{-}(t,\hbar;\xi)$ satisfies the integral 
equation
\be
\label{C2}
\frac{1}{2}\Phi^{-}(t,\hbar;\xi)+
\frac{1}{2\pi i}\int_{L}\frac{\Phi^{-}(t,\hbar;\tau)}{\tau-\xi}d\tau=\gamma(t,\hbar;\xi)
\ee
where $\gamma(t,\hbar;\lambda)$ is a principal value at infinity of the series 
$\Phi(t,\hbar;\lambda)$.

The equations (\ref{C1}),(\ref{C2}) and the condition 
$\Phi^{+}(t,\hbar;\xi)=\Phi^{-}(t,\hbar;\xi)\ast G(t,\hbar;\xi)$ imply the Fredholm integral
equation with a non-singular kernel
\be
\label{calkowe ast}
\Phi^{-}(t,\hbar;\xi)-\frac{1}{2\pi i}
\int_{L}\Phi^{-}(t,\hbar;\tau)\ast\frac{G(t,\hbar;\tau)\ast G^{-1}(t,\hbar;\xi)-1}{\tau-\xi}
d\tau=\gamma(t,\hbar;\xi)
\ee
Summarizing, the existence of a solution of the homogeneous Hilbert problem 
(\ref{zagad pierwsze}) implies that the limiting value $\Phi^{-}(t,\hbar;\xi)$ satisfies 
(\ref{calkowe ast}). The converse may not be true, as the solution of (\ref{calkowe ast}) has 
to satisfy additionally (\ref{C1}) and (\ref{C2}).

To answer under what condition the solution of (\ref{calkowe ast}) defines  a sectionally 
holomorphic solution to the homogeneous Hilbert problem consider
\beas
\Psi(t,\hbar;\lambda)=\left\{\begin{array}{lcl}
\frac{1}{2\pi i}\int_{L}\frac{\Phi^{-}(t,\hbar;\tau)}{\tau-\lambda}d\tau-\gamma(t,\hbar;\lambda)
&\mbox{for}& \lambda\in S^{+}\\
\frac{1}{2\pi i}\int_{L}\frac{\Phi^{-}(t,\hbar;\tau)\ast G(t,\hbar;\tau)}{\tau-\lambda}d\tau
&\mbox{for}& \lambda\in S^{-}\end{array}\right.
\eeas
Thus $\Psi(t,\hbar;\lambda)$ is sectionally holomorphic and vanish at infinity. The Plemelj
theorem gives
\beas
\Psi^{+}(t,\hbar;\xi)=\frac{1}{2}\Phi^{-}(t,\hbar;\xi)+
\frac{1}{2\pi i}\int_{L}\frac{\Phi^{-}(t,\hbar;\tau)}{\tau-\xi}d\tau-\gamma(t,\hbar;\xi)\\
\Psi^{-}(t,\hbar;\xi)=-\frac{1}{2}\Phi^{-}(t,\hbar;\xi)\ast G(t,\hbar;\xi)+
\frac{1}{2\pi i}\int_{L}\frac{\Phi^{+}(t,\hbar;\tau)\ast G(t,\hbar;\tau)}{\tau-\xi}d\tau
\eeas
As it is seen, from (\ref{zagad pierwsze}), the equation (\ref{C1}) is equivalent to vanishing 
of $\Psi^{-}(t,\hbar;\xi)$ on $L$. The Eq. (\ref{C2}) gives $\Psi^{+}(t,\hbar;\xi)=0$ on $L$.
Those two conditions and the fact that  $\Psi(t,\hbar;\lambda)$ is holomorphic on $S^{+}$ and 
$S^{-}$ lead to $\Psi(t,\hbar;\lambda)\equiv 0$. The integral equation (\ref{calkowe ast}) is 
a condition
\be
\label{accompanying}
\Psi^{+}(t,\hbar;\xi)=\Psi^{-}(t,\hbar;\xi)\ast G^{-1}(t,\hbar;\xi) \h\h \xi\in L
\ee
The problem of finding $\Psi(t,\hbar;\lambda)$ sectionally holomorphic, vanishing at infinity,
satisfying boundary condition (\ref{accompanying}) on $L$ is called \it  the accompanying 
problem \rm of the problem (\ref{zagad pierwsze}). Analogously, this problem implies the 
integral equation for a limit value $\Psi^{+}(t,\hbar;\xi)$
\be
\label{calkowe dolaczone ast}
\Psi^{+}(t,\hbar;\xi)+\frac{1}{2\pi i}
\int_{L}\Psi^{+}(t,\hbar;\tau)\ast\frac{G(t,\hbar;\tau)\ast G^{-1}(t,\hbar;\xi)-1}
{\tau-\xi}d\tau=0.
\ee
As it is seen from above, the solution of integral equation (\ref{calkowe ast}) defines the 
solution of original homogenous Hilbert problem (\ref{zagad pierwsze}) iff the only solution 
of the accompanying problem is the trivial one $\Psi(t,\hbar;\xi)\equiv 0$.

Thus, in order to prove the existence of the solution of the problem (\ref{zagad pierwsze})
we need to prove  that equation (\ref{calkowe dolaczone ast}) has only the trivial solution 
and that there exists a solution  of the (\ref{calkowe ast}).

To simplify the notations we will denote
\bdm
\Phi(t,\hbar;\lambda)=\sum_{m=0}^{\infty}t^{m}\Phi_{m}(\hbar;\lambda)\h\h\h
\mbox{where}\h\h\h\Phi_{m}(\hbar;\lambda)=\sum_{k=-m}^{\infty}\hbar^{k}\Phi_{m,k}(\lambda).
\edm
First observe, that the free element of 
$G(t,\hbar;\tau)\ast G^{-1}(t,\hbar;\xi)-1$
vanishes, so we can write
\bdm
G(t,\hbar;\tau)\ast G^{-1}(t,\hbar;\xi)-1=\sum_{n=1}^{\infty}t^{n}F_{n}(\hbar;\tau,\xi).
\edm
Inserting into (\ref{calkowe dolaczone ast}) one gets
\bdm
\sum_{m=0}^{\infty}t^{m}\Psi^{+}_{m}(\hbar;\xi)\, =\,  -\frac{1}{2\pi i}
\int_{L}\, \sum_{s=0}^{\infty}t^{s}\Psi^{+}_{s}(\hbar;\tau)\ast\frac{\sum_{n=1}^{\infty}
t^{n}F_{n}(\hbar;\tau,\xi)}{\tau-\xi}d\tau.
\edm
This equation can be solved iteratively
\beas
\Psi_{0}^{+}(\hbar;\xi) & = & 0\\
\Psi_{m}^{+}(\hbar;\xi) & = &  -\frac{1}{2\pi i}
\int_{L}\,\frac{\sum_{j=0}^{m-1}\Psi_{j}^{+}(\hbar;\xi)\ast F_{m-j}(\hbar;\tau,\xi)}{\tau-\xi}
d\tau
\, ,\h\h\h m\geq 1.
\eeas
Thus the only solution of (\ref{calkowe dolaczone ast}) is $\Psi^{+}(t,\hbar;\xi)=0$. 
Consequently, each solution of (\ref{calkowe ast}) defines a solution of Hilbert problem
\beas
\Phi(t,\hbar;\lambda)=\left\{
\begin{array}{l}
\frac{1}{2\pi i}\int_{L}\Phi^{-}(t,\hbar;\tau)\ast\frac{G(t,\hbar;\tau)}{\tau-\lambda}
d\tau\h\h\h\mbox{dla} \h\h\lambda\in S^{+}\\
-\frac{1}{2\pi i}
\int_{L}\frac{\Phi^{-}(t,\hbar;\tau)}{\tau-\lambda}d\tau\, +\, \gamma(t,\hbar;\lambda)
\h\h\h\mbox{dla} \h\h\lambda\in S^{-}.
\end{array}\right.
\eeas
Eq. (\ref{calkowe ast}) takes the form
\bdm
\sum_{m=0}^{\infty}t^{m}\Phi_{m}^{-}(\hbar;\xi)\, =\,  -\frac{1}{2\pi i}
\int_{L}\, \sum_{s=0}^{\infty}t^{s}\Phi_{s}^{-}(\hbar;\tau)\ast\frac{\sum_{n=1}^{\infty}
t^{n}F_{n}(\hbar;\tau,\xi)}{\tau-\xi}d\tau
+\sum_{m=0}^{\infty}t^{m}\gamma_{m}(\hbar;\xi)
\edm
and it can be solved iteratively
\bdm
\Phi_{m}^{-}(\hbar;\xi)  = \gamma_{m}(\hbar;\xi) + \frac{1}{2\pi i}
\int_{L}\, \frac{\sum_{j=0}^{m-1}\Phi_{j}^{-}(\hbar;\xi)\ast
F_{m-j}(\hbar;\tau,\xi)}{\tau-\xi}d\tau \h\h\h m=0,1,2, ...\h .
\edm
Note that the solution of (\ref{calkowe ast}), takes value in a group $\eQ$ 
iff $\gamma(t,\hbar;\lambda)\in\eQ$.

Some remarks on Riemann-Hilbert problem for $\ast$ algebra can be also found in  
(Takasaki 1994, Strachan 1997).
\paragraph{Inverse Penrose-Ward transform.}
In this paragraph we will show the correspondence between holomorphic formal bundles over 
twistor space $\PT$ and solutions of master equation (\ref{ME in H}).

As it was shown the manifold $\PT$ is covered by two coordinate neighbourhoods 
$(V,({P^{w}}, {P^{z}}, \lambda))$, $\lambda\in\mbox{\boldmath$CP$}^{1}-\{\infty\}$ and  
$(\underline{V},(\underline{P^{w}}, \underline{P^{z}}, \zeta)$,
$\zeta\in\mbox{\boldmath$CP$}^{1}-\{ 0\}$. 

Each holomorphic formal bundle over $\PT$ is characterized by a transition function
$H(t,\hbar;P^{w}, P^{z}, \lambda):V\cap\underline{V}\rightarrow\eQ$ i.e.
\bdm
H(t,\hbar;\widetilde{P^{w}}, \widetilde{P^{z}}, \lambda)=1+\sum_{m=1}^{\infty}\sum_{k=-m}^{\infty}
t^{m}\hbar^{k}H_{m,k}(P^{w}, P^{z}, \lambda)
\edm
with $H_{m,k}(P^{w}, P^{z}, \lambda)$ being holomorphic. The pull back of the series by the
map $p:\M\times\mbox{\boldmath$CP$}^{1}\rightarrow\PT$ gives on the correspondence space
${\cal F}=\M\times\mbox{\boldmath$CP$}^{1}$ the series ${\cal H}=p^{\ast}H$ constant along 
each twistor surface
\be
\label{stale H na pow_twist}
\ell_{w}{\cal H}=0\, ,\h\h\h \ell_{z}{\cal H}=0.
\ee
Briefly we will write 
${\cal H}(t,\hbar;\lambda)\equiv {\cal H}(t,\hbar;w,z,\tilde{w},\tilde{z},\lambda)$.

\vspace{1 ex}

As $H$ is a transition function, ${\cal H}(t,\hbar;\lambda)$ can be factorized
\be
\label{factor 1}
\Psi(t,\hbar;\lambda)=\underline{\Psi}(t,\hbar;\lambda)\ast{\cal H}(t,\hbar;\lambda)
\h\h\h\mbox{for}\h\h\h\lambda\in \mbox{\boldmath$CP$}^{1}-\{ 0,\infty\}.
\ee
where  $\Psi(t,\hbar;\lambda)$ is holomorphic everywhere apart from $\lambda=\infty$ and
$\underline{\Psi}(t,\hbar;\lambda)$ is holomorphic everywhere apart from  $\lambda=0$ and 
both series take values in $\eQ$. 

The problem of such factorization, known as Riemann-Hilbert problem, reduces to the 
previously discussed homogenous Hilbert problem. 

Indeed, let $L$ be a smooth contour on $\mbox{\boldmath$CP$}^{1}$ (for example an equator).
One can find $\Psi(t,\hbar;\lambda)$ holomorphic on $S^{+}$ (we use the same notation as in 
the previous paragraphs), $\underline{\Psi}(t,\hbar;\lambda)$ holomorphic on $S^{-}$ and 
continous on $S^{+}\cup L$ and $S^{-}\cup L$, respectively. On $L$ they satisfy the 
condition
\bdm
\widetilde{\Psi}^{+}(t,\hbar;\xi)=\underline{\Psi}^{-}(t,\hbar;\xi)\ast
{\cal H}(t,\hbar;\xi)\h\h\h\h\xi\in L.
\edm
The series $\Psi$ and $\underline{\Psi}$ can be analitically continued onto 
$S^{-}-\{\infty\}$ and $S^{+}-\{0\}$, respectively, by
\bea
\label{3.27}
\begin{array}{lcl}
\mbox{for}\h\h\lambda\in S^{-}-\{\infty\}& & \Psi(t,\hbar;\lambda):=
\underline{\Psi}(t,\hbar;\lambda)\ast {\cal H}(t,\hbar;\lambda)\\
\mbox{for}\h\h\lambda\in S^{+}-\{ 0 \} & & \underline{\Psi}(t,\hbar;\lambda):=
\Psi(t,\hbar;\lambda)\ast {\cal H}^{-1}(t,\hbar;\lambda)
\end{array}
\eea
In this way we obtain $\Psi(t,\hbar;\lambda)$ defined in each finite point of the comlex 
plane and sectionally holomorphic on $S^{+}$, $S^{-}$, satisfying on $L$ the condition 
$\Psi^{+}(t,\hbar;\xi)=\Psi^{-}(t,\hbar;\xi)$. This means that such a 
$\Psi(t,\hbar;\lambda)$ is holomorphic on the whole complex plane, as desired.
Analogously $\underline{\Psi}(t,\hbar;\lambda)$ is holomorphic on 
$\mbox{\boldmath$C$}-\{0\}$. From the definition (\ref{3.27}) the factorization 
(\ref{factor 1}) holds.

\vspace{1 ex}

Suppose we are given $\Psi(t,\hbar;\lambda)$ and $\underline{\Psi}(t,\hbar;\lambda)$
defined by (\ref{factor 1}). Thus from (\ref{stale H na pow_twist}) one gets
\bdm
\ell_{\alpha}[\underline{\Psi}^{-1}(t,\hbar;\lambda)
\ast\Psi(t,\hbar;\lambda)]=0\, , \h\h\h\alpha=w,z.
\edm
and 
\bdm
\ell_{\alpha}\Psi(t,\hbar;\lambda)\ast
\Psi^{-1}(t,\hbar;\lambda)=
\ell_{\alpha}\underline{\Psi}(t,\hbar;\lambda)
\ast\underline{\Psi}^{-1}(t,\hbar;\lambda).
\edm
The LHS is holomorphic everywhere apart from $\lambda=\infty$, while RHS is holomorphic 
everywhere apart $\lambda=0$ and at infinity it may have only a first order pole. Thus 
from the \it Liouville theorem \rm they both are linear with respect to $\lambda$ i.e. 
\beas
\ell_{\alpha}\Psi(t,\hbar;\lambda)&=&
\frac{1}{i\hbar}
(-A_{\alpha}+\lambda \epsilon_{\alpha\beta}Gg^{\tilde{\sigma}\beta}
A_{\tilde{\sigma}})\ast\Psi(t,\hbar;\lambda)\\
\ell_{\alpha}\underline{\Psi}(t,\hbar;\lambda)&=&
\frac{1}{i\hbar}
(-A_{\alpha}+\lambda \epsilon_{\alpha\beta}Gg^{\tilde{\sigma}\beta}
A_{\tilde{\sigma}})\ast\underline{\Psi}(t,\hbar;\lambda)
\eeas
where $A_{\alpha}=A_{\alpha}(t,\hbar)$, $\alpha= w,z$, 
$A_{\tilde{\sigma}}=A_{\tilde{\sigma}}(t,\hbar)$, $\tilde{\sigma}=,\tilde{w},\tilde{z}$
do not depend on $\lambda$. 

The $A_{\alpha}$, $A_{\tilde{\sigma}}$ satisfy the SDYM equations
(\ref{SDYM 1}), (\ref{SDYM 2}) and (\ref{SDYM 3}). This is easily seen from the fact that the 
vector fields $\ell_{w},\ell_{z}$ commute for each value of $\lambda$. Thus
\beas
\lefteqn{0=\epsilon^{\beta\alpha}\ell_{\beta}\ell_{\alpha}\Psi(\lambda)=
\epsilon^{\beta\alpha}(\partial_{\beta}-\lambda G\epsilon_{\beta\delta}g^{\tilde{\sigma}\delta}
\partial_{\tilde{\sigma}})[(\frac{-1}{i\hbar}A_{\alpha}+\frac{1}{i\hbar}\lambda 
G\epsilon_{\alpha\sigma}g^{\tilde{\rho}\sigma}A_{\tilde{\rho}})\ast\Psi]=}\\
&=&\frac{1}{i\hbar}[-\epsilon^{\beta\alpha}\partial_{\beta}A_{\alpha}+
\frac{1}{i\hbar}\epsilon^{\beta\alpha}A_{\alpha}\ast A_{\beta}]+
\frac{\,\lambda\, }{i\hbar} Gg^{\tilde{\sigma}\alpha}[\partial_{\tilde{\sigma}}A_{\alpha}-
\partial_{\alpha}A_{\tilde{\sigma}}+\frac{1}{i\hbar}(A_{\tilde{\sigma}}\ast A_{\alpha}-
A_{\alpha}\ast A_{\tilde{\sigma}})]+\\
& &+\frac{\lambda^{2}}{i\hbar}\, \frac{G}{\tilde{G}}\epsilon^{\tilde{\rho}\tilde{\sigma}}
[\partial_{\tilde{\rho}}A_{\tilde{\sigma}}+\frac{1}{i\hbar}A_{\tilde{\rho}}\ast 
A_{\tilde{\sigma}}].
\eeas
Apropriate terms at $\lambda^{0}, \lambda^{1}, \lambda^{2}$
gives SDYM: (\ref{SDYM 1}), (\ref{SDYM 3}) i (\ref{SDYM 2}), respectively.

Let us note that the factorization (\ref{factor 1}) does not define the series 
$\Psi(t,\hbar;\lambda)$, $\underline{\Psi}(t,\hbar;\lambda)$ uniquely. They can be 
symultaneously multiplied by $a(t,\hbar)\in\eQ$, independent of $\lambda$.
This implies the gauge freedom for SDYM potential.

Thus we can choose such $\Psi(t,\hbar;\lambda)$, $\underline{\Psi}(t,\hbar;\lambda)$ 
that $A_{\alpha}=0$. Then (compare with (\ref{Laxa}))
\be
\label{def rozw rown ME}
\ell_{\alpha}\Psi(t,\hbar;\lambda)\ast
\Psi^{-1}(t,\hbar;\lambda)=\frac{1}{i\hbar}\partial_{\alpha}\Theta
\ee
where $\Theta$ is a solution of master equation (\ref{ME in H}).
b\rm .

\section{Conclusions}
In this work we have found the evidence of integrability of $\ast$-SDYM equations. 
This evidence follows from:

\bi
\item the existence of infinite number of conservation laws,
\item the existence of Lax pair,
\item the one-to-one correspondence between solutions of $\ast$-SDYM equations and 
formal holomorphic bundles over $\PT$ with structure group $\eQ$. 
\item the existence of solution to the Riemann-Hilbert problem what gives rise 
to an algebraic method of generating solutions to (ME).
\ei
In the second part of this paper some examples of reductions of 
$\ast$-SDYM to other integrable systems such as $SU(N)$-SDYM equations,
$SU(N)$ chiral equations and heavenly equations will be given. We also find  
a sequence of $SU(N)$ chiral fields tending to the heavenly space when
$N\rightarrow\infty$.

\section*{Acknowledgments}
We are grateful to Maciej Dunajski and Jacek Tafel for valuable discussions. 
The work was partially supported by CONACyT grant 41993-F (Mexico) and NATO grant 
PST.CLG. 978984.


\begin{thebibliography}{  }
\bibitem{} Asakawa T. and Kishimoto I. (2000), \it Noncommutative Gauge Theories from 
Deformation Quantization\rm , hep-th/0002138
\bibitem{} Bayen F., Flato M., Fronsdal C., Lichnerowicz A. and Strenhaimer D. (1978), \it Ann. Phys. 
\rm\bf 111\rm , 61-110 and 111-150.

\bibitem{} Birkhoff G.D. (1913), \it Math. Annalen \bf 74\rm , 122-133.

\bibitem{} Boyer C.P. and Plebański J.F. (1985), \it J. Math. Phys. \bf 26 \rm (2) 229-234.

\bibitem{} Brezin E., Itzykson C., Zinn-Justin J. and Zuber J.B. (1979), \it Phys. Lett. \rm\bf B82 \rm 442.

\bibitem{} Chau L-L. (1983), \it Chiral Fields, SDYM Fields as Integrable Systems,
and the Role of the Kac-Moody Algebra\rm , in: \it Nonlinear phenomena\rm , ed. K.B. Wolf,
Lecture notes in Physics \bf 189 \rm (Springer, New York) 110-127.

\bibitem{} De Wilde M. and Lecomte P.B.A. (1983), \it Lett. Math.
Phys. \bf 7\rm , 487- .

\bibitem{} Donaldson S.K. (1985), \it Proc. Lond. Math. Soc. \bf 3\rm , 1-26.

\bibitem{} Dunajski M. and Mason L.J. (2000), \it Comm. Math. Physics \bf 213 \rm , 641-672.

\bibitem{} Husain V. (1994), \it Class. Quantum Grav. \rm\bf 11\rm , 927-937.

\bibitem{} Fairlie D.B., Fletcher P. and Zachos C.K. (1990), \it J. Math. Phys. \rm\bf 31 \rm (5), 1088-1094\\
Fairlie D.B. and Zachos C.K., \it Phys. Lett. \rm\bf B 224 \rm no. 1,2, 101-107 (1989)\\
Fairlie D.B., Fletcher P. and Zachos C.K., \it Phys. Lett. \rm\bf B 218 \rm no. 2, 203-206
(1989).

\bibitem{} Fedosov B.V. (1994), \it J. Diff. Geom\rm . \bf 40\rm , 213-238.

\bibitem{} Fedosov B.V. (1996), \it Deformation Quantization and Index Theory \rm (Akademie Verlag, Berlin).

\bibitem{} Flaherty E.J. (1976), \it Hermitian and K\H{a}hlerian
Geometry in Relativity\rm , Lecture Notes in Physics (Springer-Verlag, Berlin).

\bibitem{} Forma\'{n}ski S. (2004), Moyal deformation of heavenly equations. 
Integrability and relations to other non-linear equations\rm . PhD thesis Tech. Univ. of Lodz, 
2004 (in Polish)

\bibitem{} Jacobson N. (1980), \it Basic Algebra II \rm (Freeman, San Francisco, CA)

\bibitem{} Ko M., Ludvigsen M. and Newman E.T. and Tod K.P. (1981),
\it Phys. Reports \rm\bf 71\rm , No.2, 51-139.

\bibitem{} Kontsevich M. (1997), \it Deformation quantization
of Poisson manifolds\rm , q-alg/9709040.

\bibitem{} Kupershmidt B.A. (1990), \it Lett. Math. Phys. \bf 20\rm , 19-31.

\bibitem{} Leznov A.N. (1988), \it Theor. M. Phys. \bf 73\rm , 1233-1237

\bibitem{} MacLane S.(1939), \it Bull. Amer. Math. Soc. \bf 45\rm , 888-890.

\bibitem{} Mason L.J. and Newman E.T. (1989), \it
Comm. Math. Phys. \rm\bf 121  \rm 659-668.

\bibitem{} Mason L.J. (1990), \it Twistor Newsletter\rm , \bf 30\rm , 14-17.

\bibitem{} Mason L.J. and Woodhouse N.M.J. (1996), \it Integrability,
self-duality, and twistir theory \rm (Clarendon Press, Oxford).

\bibitem{} Muscheliszwili N.I. (1962), \it Singular integral equations \rm (Moskwa)

\bibitem{} Nair V.P. and Schiff J. (1990), \it Phys. Lett. \bf B 246\rm , 423.

\bibitem{} Neumann B.H. (1949), \it Trans. Amer. Math. Soc. \bf 66\rm , 202- .

\bibitem{} Newman E.T. (1978), \it Phys. Rev. \rm\bf D18\rm ,2901-2908.
 
\bibitem{} Parkes A. (1992), \it Phys. Lett. \bf B 286\rm , 265.

\bibitem{} Park Q-H. (1990), \it Phys. Lett. \rm\bf B 238\rm ,
No. 2,3,4, 287-290.

\bibitem{} Park Q-H. (1992), \it Int. J. Mod. Phys. \rm\bf A 7\rm ,
No. 7, 1415-1447.

\bibitem{} Penrose R. (1976), \it Gen. Rel. Grav. \bf 7\rm , No. 1, 31-52.

\bibitem{} Penrose R. and Ward R.S. (1980), \it Twistors
for flat and curved space-time \rm  in: \it General Relativity and Gravitation \rm
ed. A. Held, Vol. 2, (New York, London: Plenum Press), 283-328.

\bibitem{} Pleba\'{n}ski J.F. (1975), \it J. Math. Phys. \bf 16\rm (12) 2395-2402.

\bibitem{} Pleba\'{n}ski J.F. and Hacyan S. (1975),
\it J. Math. Phys. \bf 16 \rm (12)  2403-2407.

\bibitem{} Pleba\'{n}ski J.F. and Przanowski M. (1996), \it Phys. Lett. \bf A 212\rm , 22.

\bibitem{} Plemelj J. (1908), \it Monatsheft f\H{u}r Math. und Phys. \bf
19\rm , 211-245.

\bibitem{} Pogorzelski W.(1966), \it Integral equations and their applications. \rm
Pergamon Press PWN, Oxford.

\bibitem{} Prasad M.K., Sinha A. and Chau Wang L-L.(1979), \it Phys.
Lett. \bf B 87\rm , 237.

\bibitem{} Przanowski M. and Forma\'{n}ski S. (1999), \it Acta Phys. Pol. \rm \bf B 30 \rm , 863-879.

\bibitem{} Przanowski M., Plebański J.F. and Formański S.(2001 a),
\it Integrability of SDYM Equations for the Moyal bracket Lie algebra\rm , w tomie: \it Exact
Solutions and Scalar Fields in Gravity\rm , eds. A. Mac'initials, J.L. Cervantes-Cota and C.
L\H{a}mmerzhal (Kluwer Academic Publishers, USA).

\bibitem{}  Przanowski M., Plebański J.F.  and Formański S. (2001 b),
\it Deformation Quantization of $SDiff(\Sigma^{2})$ SDYM Equation\rm , in: \it Developments
in Mathematical and Experimental Physics\rm , Vol. A, eds. A. Macias, F. Uribe and E. Diaz
(Kluwer Academic, Plenum Publishers).

\bibitem{} Ruiz J.M. (1993), \it The Basic Theory of Power Series\rm , (Braunschweig: Vieweg-Verlag)

\bibitem{} Strachan I.A.B. (1992), \it Phys. Lett. \rm\bf B 282\rm , 63-66.

\bibitem{} Strachan I.A.B. (1993), \it Class. Quantum Grav. \rm \bf 10\rm , 1417-1423.

\bibitem{} Strachan I.A.B. (1997), \it J. Geom. Phys. \rm\bf 21\rm , 255-278.

\bibitem{} Takasaki K. (1990), \it J. Math. Phys. \rm\bf 31 \rm (8), 1877-1888.

\bibitem{} Takasaki K. (1994), \it J. Geom. Phys. \rm\bf 14 \rm , 111-120 and 332-.

\bibitem{} Ward R.S. (1985), \it Integrable systems in twistor theory, \rm in:
\it Twistors in mathematics and physics, \rm eds. T.N. Bailey and R.J. Baston

\bibitem{} Ward R.S and Wells R.O. (1990), \it Twistor geometry and field theory., \rm 
Cambridge University Press, Cambridge.

\bibitem{} Yang C.N. (1977), Phys. Rev. Lett. \bf 38\rm , 1377-1379.
\end{thebibliography}
\end{document}